# On modeling energetic electrons in laser fusion plasmas

Wallace Manheimer

Retired from the US Naval Research Lab (NRL)

June, 2026

Abstract

It has been known for decades that laser fusion can be and often is plagued by the production of energetic electrons, produced either by an instability, or by the energetic tail of a Maxwellian distribution. Despite more than 25 years of a variety of efforts, there is still no generally accepted way treat or address this issue. This work hopes to advance the progress by developing a Fokker Planck model which is simple enough to be used at every time step of a rad-hydro simulation, and accurate enough to be useful. It makes several approximations, the main one being that the percentage of these energetic electrons is small. Recent experiments confirm that so far, this has proven to be the case for plasmas subject to laser plasma instabilities. Hence energetic electrons interact with the background plasma and not with each other. This work makes no attempt to solve for the entire electron distribution function. It makes a variety of other approximations to solve the resulting Fokker Planck equation. This rather lengthy report both summarizes the author's work with the NRL group and presents his new advances since the termination of the NRL project. An important result is a tentative conclusion that energetic electrons may not be such a big problem for laser fusion.

Key words: laser fusion, energetic electrons, energetic electrons for laser fusion, Fokker Planck equation, solutions to the Fokker Planck equation

## I. Introduction

Energetic electrons in a laser produced plasma cannot always be described with a local theory. That is the mean free path of some electrons may be comparable to or even greater than the temperature gradient length. There are at least two sources of these energetic electrons. First there can be laser plasma instabilities which generate many very energetic electrons. Second the tail of the Maxwellian distribution may also decouple locally. This work examines both.

It has been 26 years since Schurtz et al (1) have published their work on nonlocal transport of energetic electrons in laser fusion targets. In their case, the electrons they were looking at were in the tail of the Maxwellian. Yet there is still no generally accepted way to treat this important issue, illustrating both the difficult nature and importance nature of this problem. What is required is a theory that is both accurate enough and is simple enough that it can be applied at every time step, or nearly every time step in a rad hydro code.

The theory in Ref (1) is basically a Krook theory and other groups have developed similar versions (2-8). At NRL we also developed Krook models (9-14). One problem is that when these theories were applied to direct drive laser fusion, the calculations at least at NRL (15), the University of Rochester Laboratory for Laser Energetics, URLLE (16), and possibly at other labs, showed that there was no gain. To this author's knowledge, none of these negative results had ever been published.

A little analysis of the Krook model shows that it will overestimate the preheat. Consider what the model used. A particle stream moves through the background particles which it collides with. After it travels one mean free path, 1/e of these particles remain, those that don't remain have fallen into the background. The next mean free path, same thing and so on. There is no slowing down of the particles. Hence the population of energetic particles decrease exponentially, but rather slowly as they travel.

The Fokker Planck model which we use here is characterized by both friction and diffusion. Let us consider the friction. The friction force is a rapidly decreasing function of particle energy, so as the particle slows down, it friction force gets stronger.

For the case at hand, the particle equation as it moves along a straight-line orbit is:

$$V\frac{dV}{dx} = -\frac{\nu(V_o)}{V^2}V_o^3. \qquad (1)$$

Here $V_o$ is the particle's initial velocity. Equation (1) has a simple solution:

$$V = [V_o^4 - 4\nu(V_o)V_o^3 x]^{1/4} \qquad (2)$$

Hence the particle cannot travel even a single mean free path, evaluated at its initial velocity, before it stops. Including this effect is the basic reason that one would think that accounting for it would reduce the fuel preheat.

This was hardly ignored by scientists studying the effect of energetic electrons. For instance, Ref (6) put what we have called an iron dome on top of the deposition profile calculated by the Krook analysis. Reference (8) came up with a modified Krook model. Instead of using $f-f_m$ on the rhs, it used derivatives. Reference (17) devised a Krook model but added to it a downward energy cascade to mockup the effect of the dynamical friction. Others modified the multigroup diffusion model (18,19), usually used as an approximation to describe neutron transport, to model energetic electron transport. All of these are reasonable efforts to deal with what is a fundamental error of the Krook model. However, this author believes we can do better, by starting out with a correct model. He thinks the first step on the ladder is make approximations to this model so as to see if works for the simplest problems. If this works out, then make better approximations to see if it would work in more complicated problems. At this point, the author hopes to be able to claim that we have at least achieved the first step.

Some researchers have decided to go a different route and come up with what one would call a full simulation (20-22). These calculate many other effects, magnetic fields, electron inertia, this or that instability, Hall effects… These codes are much more complicated; there is little chance that they could be used at each step of a rad hydro simulation, as the authors in fact stipulate (20) (and if they were, there would be little need for the rad-hydro simulation). Furthermore, they seem to be more sensitive and must put in additional guard rails to keep them from crashing.

Mea culpa, the author admits that at the NRL, we were probably the last group to realize that Krook models were not viable. However instead of finding ways to patch it up, we abandoned it completely (well, almost) and went directly to a Fokker Planck equation (23-26). We made several simplifications to get results. We certainly were not looking at the large variety of effects of say (22) but were attempting to evaluate the effect of the energetic electrons on the implosion. Many of these results seemed to be physically reasonable, they matched reasonably well experiments on the National Ignition Facility/ Lawrence Livermore National Lab (NIF/URLLE) study of energetic electrons generated by the Absolute Raman side scatter instability (27). Also at least we started with something which was basically correct and pushed it as far as we could. As improvements become necessary, they could likely be made by extending the model developed here, rather than adding another correction to a different model which is fundamentally wrong. In fact, this paper gives some calculations made by improving and extending the work in the initially published results. It also suggests additional improvements that could be made, especially for the case of the energetic electrons generated in the tail of the Maxwellian.

The author must admit that for the past few years he has been very excited and motivated by the recent triumph at LLNL, where they demonstrated that a burning plasma is possible (28,29,30), probably at least 15-20 years before ITER can make the same claim. Furthermore, the cost of the NIF effort is quite small compared to the cost to ITER, and its tritium demand for the long research stage is much less. NIF is a frequency tripled glass laser, having a wavelength of 335 nm and producing as much as 2 MJ of laser light energy. The key is that the alpha particle has much lower energy (3.5 MeV) than the neutron (14 MeV). Therefore, if the plasma configuration is produced just right, the alpha can be absorbed locally while the neutron escapes. Hence the alpha can start a burn wave, and the plasma can heat, at least for a while, as it expands. This is what the LLNL group has succeeded in demonstrating. In other words, the laser is like the sparkplug in the cylinder of a car. It does not provide the energy to drive, but it just ignites the fuel which does power the car. The author has gone record as saying that 100 years from now, this will probably be regarded as one of the main experiments of 21st century physics, and likely will have won a Nobel prize (31, 32).

Furthermore, this work has immediately solved another of the most vexing problems in magnetic fusion, namely what to do with the alpha particles. In an ideal magnetic fusion world, the alphas would heat the plasma, just the right amount, and then disappear. But how do they do this? Any electromagnetic way of getting rid of the alphas would probably also get rid of the deuterium, since they have the same e/M.

The author sees laser fusion as suddenly coming up with 2 gigantic achievements, first producing a likely scalable device with Q>1, and then solving the alpha dilemma. Hence for these two reasons, the author has gone on record of advising the DoE to put its main emphasis for fusion for energy (as opposed to stockpile stewardship), on laser fusion instead of magnetic fusion (31, 32). An image of LLNL's configuration from (28) is in Figure (1).

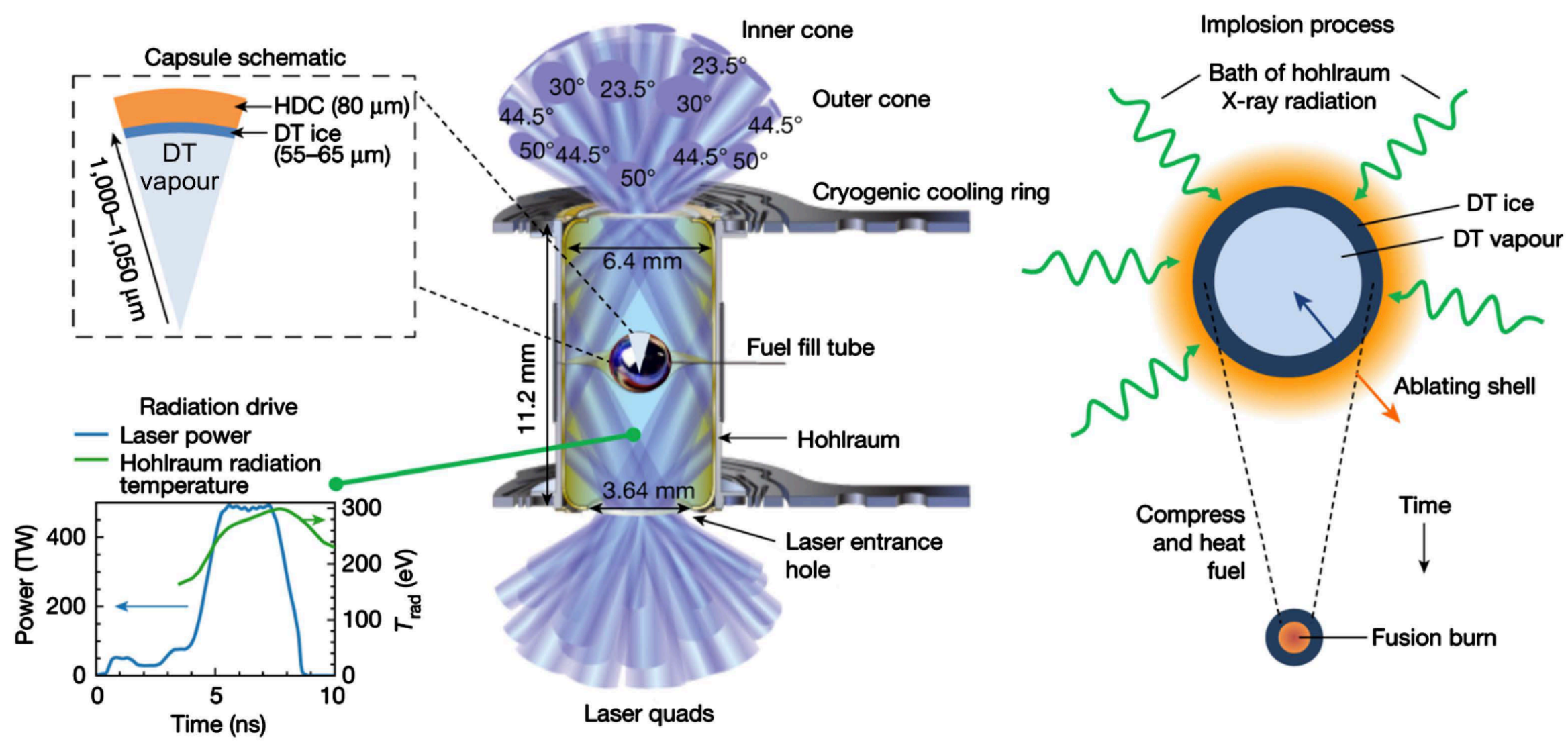

Figure 1.  The configuration of the LLNL successful experiment which produced an alpha particle burn wave.  The target is compressed by X-Rays produced on the wall of the chamber, called a hohlraum, containing the target.

The experiment in 2021 achieved a gain just below unity, but showed that they produced an alpha burn wave by showing that the neutron production maximized *after* the bang time.   A graph taken from Ref (28) is reproduced in Fig. (2).  Other diagnostics actually showed the plasma initially heating as it expanded (30).  Recently the group has published a review paper of their recent work (29).  This work showed that they now have achieved a gain of ~ 2.5, and it predicts that future higher gains are achievable with NIF.

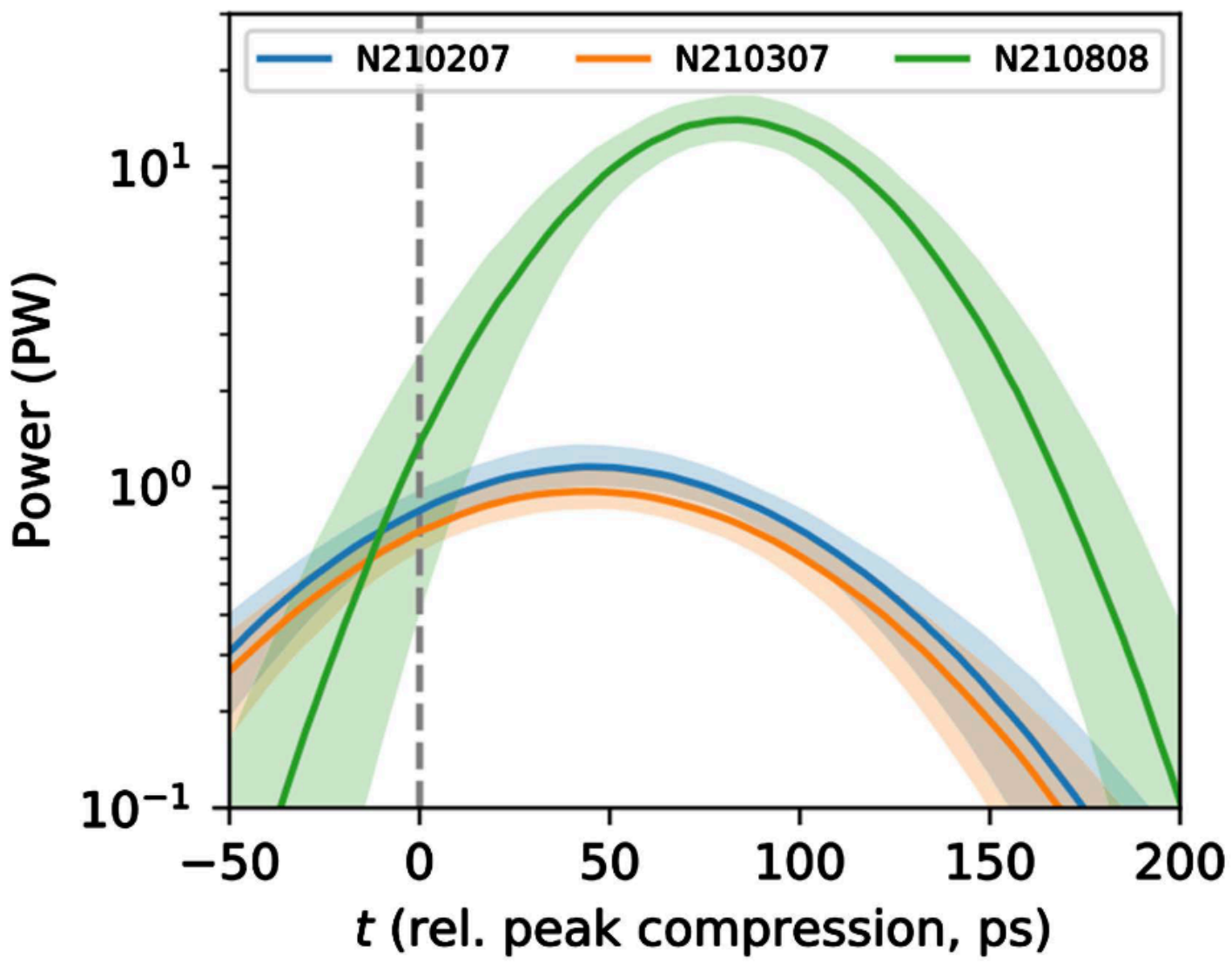

Figure 2.  A plot of fusion power at the times, in picoseconds, around the bang time of the implosion.  Notice that the power maximizes *after* the bang time, meaning that an alpha burn wave has been produced.

The LLNL work is not supported for energy, but for nuclear stockpile stewardship.  The configuration they have used, is a target that is centered in a heavy metal hohlraum.  The laser illuminates the inner hohlraum wall.  A dense, hot, high Z plasma is formed, and the black body X-Rays emitted illuminate and compress the target.  This configuration is called indirect drive. However, the LLNL group estimates that only 10-15% of the laser energy strikes the target. Since the implosion is caused by X-Rays, and not by ultraviolet light, it is unlikely that energetic electrons will be a factor at the target, although they may be important near the hohlraum. Indirect drive may or may not be a viable configuration for energy production, but it is certainly what the sponsors of stockpile stewardship insist on, as this approach most closely approximate the physical processes going on in a nuclear weapon.

The group at URLLE has also made important gains.  An alternate configuration, called direct drive, is to have the laser directly illuminate the target. URLLE has an ultraviolet (wavelength =

335nm) laser called OMEGA. It is also a gigantic laser (but tiny compared to NIF) which produced ~ 30kJ of laser light. With direct drive, perhaps as much as 100% of the laser energy is available to strike the target. This has been the approach at NRL, and mostly also at URLLE. Other researchers have expressed skepticism that direct drive could ever work. They thought that only X-rays, whether in a hohlraum or nuclear weapon, have the necessary qualities to generate the required implosions. In fact, so far, it has been the only way this has been done. The skeptics think that ultraviolet light is unlikely to be able to this (33).

Here is where URLLE made a very important advance. Their laser does not have the energy required to generate an implosion which can produce a plasma large enough to locally absorb the alpha particles, so they are unable to generate the alpha burn wave that NIF was able to produce. Their effort was to produce what they called implosions hydrodynamically equivalent to what NIF was able to do. That is, to produce an implosion just like NIF's but on a much smaller length scale, and which would come apart much more quickly. In their central hot spot, they would produce neutrons at the same rate as NIF, but of course they would have no capability of producing nearly as many, and certainly could not produce an alpha particle burn wave.

The URLLE effort was no simple job. In 2013, attempting to compress cryogenic spheres with OMEGA they could produce only about $\sim 10^{13}$ neutrons (34). Their hydrodynamically equivalent implosion would produce more like $\sim 3 \times 10^{14}$ fusion neutrons; it took a tremendous effort on their part, but they finally achieved this (35,36). This is as opposed to NIF's $\sim 2 \times 10^{18}$, exploiting an alpha burn wave. To see the importance of the alpha burn wave, note that OMEGA, with 30 kJ produced $\sim 3 \times 10^{14}$, while NIF, with about 60 times more energy, produced about 6000 times more neutrons (29).

Figure (3) contains two plots. The left is URLLE's resulting neutron production from 2013, with their 1D calculations predicting that they should be able to produce more than $10^{14}$. With a ten-year effort on both theory and experiment, they were finally able to produce the hydrodynamically equivalent implosion as shown in the right plot.

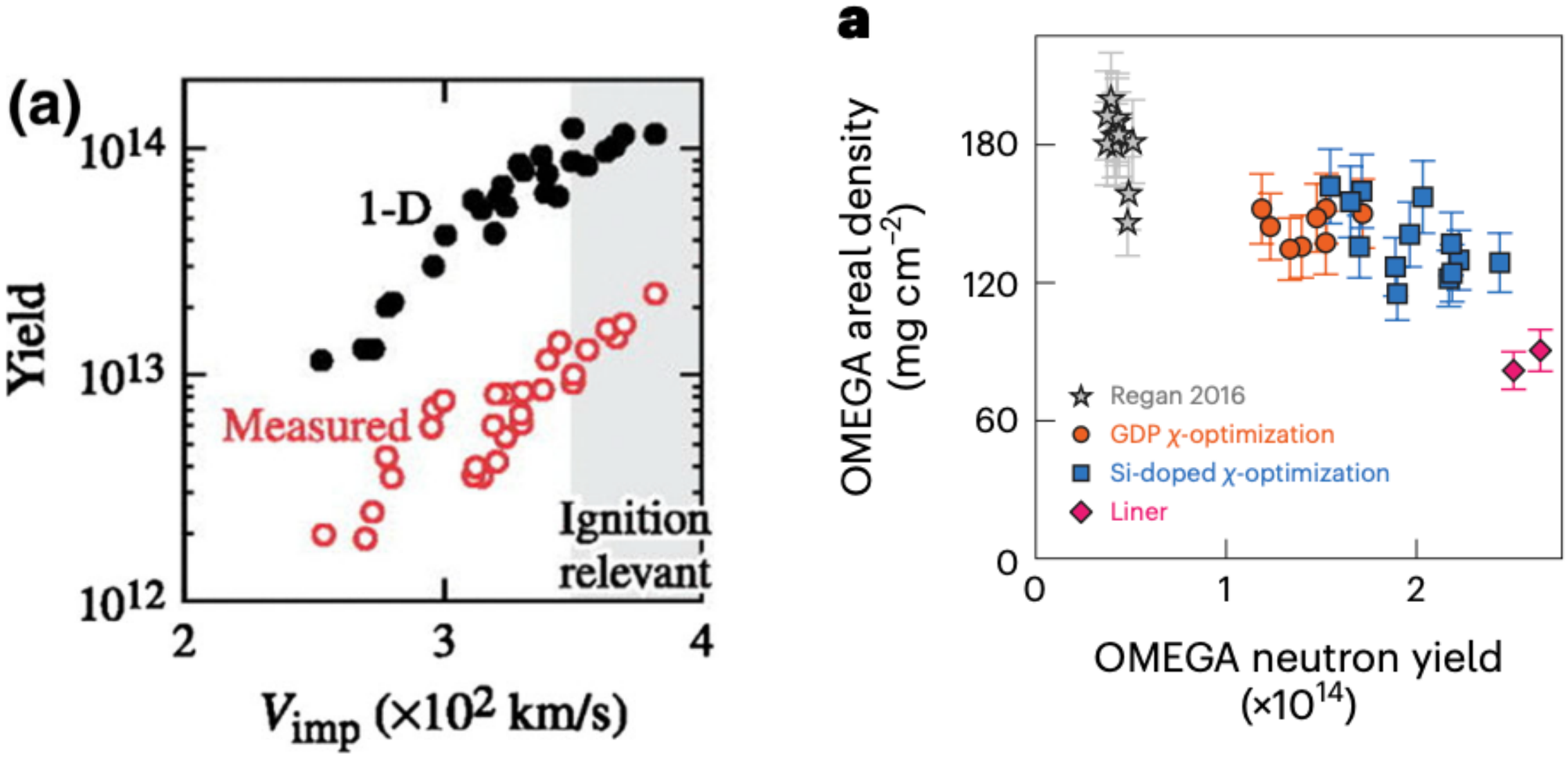

Figure (3) Left is a plot of neutrons emitted from the target in their 2013 effort. Notice that their code result predicted that they would get just over $10^{14}$ neutrons, but they only got ~ $2x10^{13}$. On the right is their result from 2024 after tremendous effort on both the theory and experiment. They increased the neutron production to ~ $3x10^{14}$, more than an order of magnitude increase. This is just about what they expected from a hydrodynamically equivalent implosion.

This author enthusiastically congratulates the LLNL and URLLE groups on starting out with an initial failure, which discouraged many (including the author), and ending ten years later with tremendous successes.

While the URLLE results certainly provide encouragement to the direct drive, one can reasonably ask whether they are missing anything. Is there some failure mode, hiding out of sight? To this author, the obvious thing the URLLE results do not seem to address, is the issue of energetic electrons. Because the gradient scale length is so small, it is almost a sure thing that main instabilities are stabilized. Not only is the length scale so small, but so is the time scale. The tails of the Maxwellian are almost always filled by diffusion from the body electrons. However, as the tail energy increases, the diffusion time to fill up the tail from the body at energy Ξ, is proportional to Ξ/T where T is the electron temperature. To briefly summarize, the short time and short scale length of the URLLE implosions may prevent energetic electrons from forming. It is unlikely that experiments with NIF sized lasers will have this luxury. This is one reason the author has been so motivated to continue to examine this issue, even though the NRL program (which paid his salary) had terminated a few years ago. While it is way too early to come to any conclusions, the author's work at this point indicates that the energetic electrons may well do less harm than initially thought.

The other lab which made significant contributions, is my own, NRL.  For decades, this lab has developed a new laser concept to apply to laser fusion.  This is the use of excimer lasers as a driver.  Originally the concept was KrF lasers, and decades ago, NRL built the NIKE laser.  It had a wavelength of 248nm and produced a very smooth optical beam with an energy of ~ 2kJ.  Also, it built a rep rated laser ELECTRA, with a pulse energy of ~ 200J.  The author believes it is still the highest *average* power laser relevant for laser fusion.  Since the active material is a fast-flowing gas, there is no laser glass that heats up and must cool down before the net shot.  Hence it is likely that excimer lasers have better capability for average power than the glass lasers used in the other labs.  Obviously, NIKE and ELECTRA are flea powered compared to the lasers at URLLE and NIF, but by examining smaller area targets, both internal and external parts, the lab was able to do important work. (37)

More recently the interest at the lab has shifted to Argon Fluoride laser with a wavelength of 193nm.  Virtually all theories predict that shorter wavelength is advantageous for laser fusion.  Figure (4) shows NRL calculations of fusion gain for frequency tripled glass lasers, KrF laser, and ArF lasers (38).  It has been argued that ArF lasers have many more of the requirements for laser fusion than do glass lasers (39).  Just before the NRL laser fusion was terminated, the lab transitioned ELECTRA to ArF and showed that it had about the same parameters as when it ran on KrF.  Also shown in Fig (4) is the setting up of ELECTRA for ArF use.

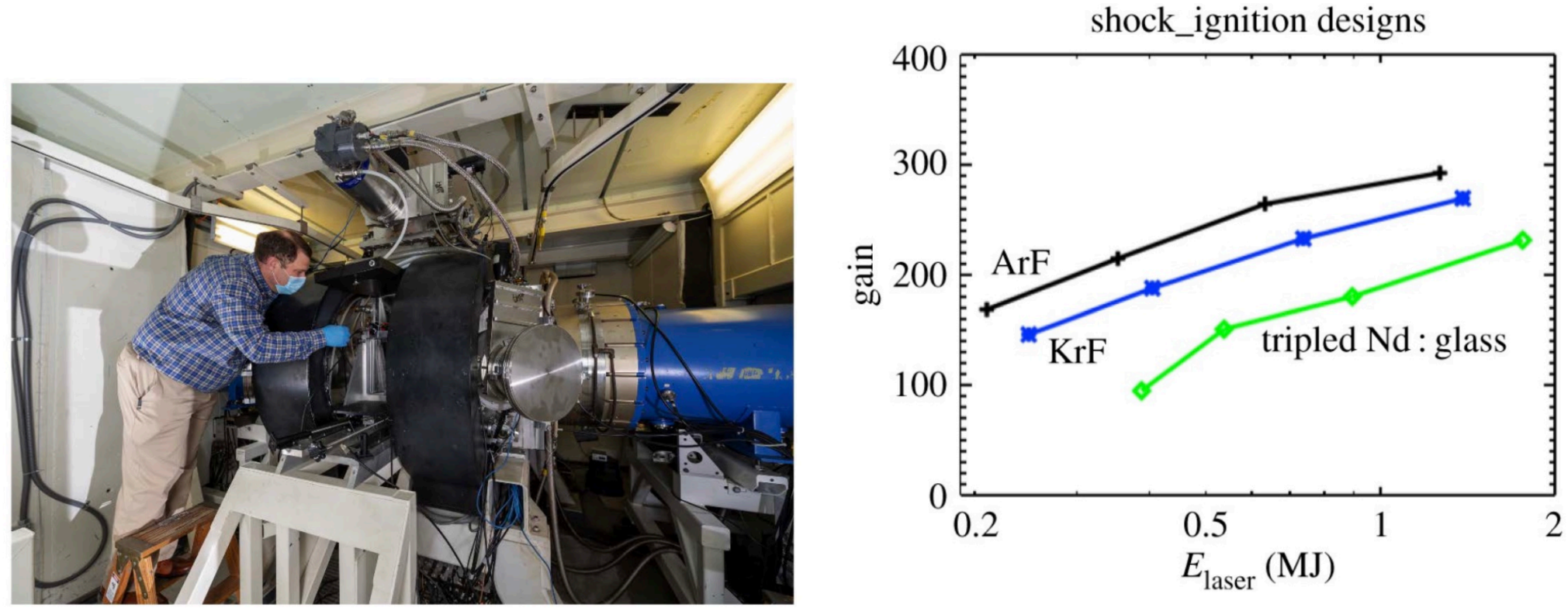

Figure (4)  Left:  Setting up the NRL laser ELECTRA for argon fluoride.  Right NRL calculations of gain versus laser energy for 3 different lasers.  According to these calculations, ArF has the potential of roughly doubling the gain of frequency tripled glass lasers.

Consider the graph of laser Q  in Fig (4).  It shows gains as high as 250.  References (31, 38, and 39) also assert that the efficiency of the laser could be as high as 10%.  Let's consider for a moment what this means for energy production.  Say laser produces produce ~1 Joule of laser light energy.  This produces ~250 Joules of fusion energy.  However, for such a new power source, let us say this generates electricity classically with an efficiency of ~1/3.  In other words, it produces ~80 Joules of electric energy. But the 10% efficient laser demands ~10 Joules, so ~70 Joules is left over for the grid. It seems like this could be a viable energy source for the civilian economy.

The paper is organized as follows. Section II reviews results of LPI experiments in laser fusion configurations. Section III goes over a derivation of the Fokker Planck Equation for deposition of energetic electrons in laser fusion targets. Sections IV and V discuss solutions of the Fokker Planck equation by what we call the method of characteristics. In Section IV, its approximation is to throw out some of the smaller terms in the equation and come up with exact solutions. Section V shows how one might approximate the result while including the neglected terms. Sections VI and VII discuss another method of solution which we call sparse eigenfunctions. In Sections VI all terms in the equation are retained, and one finds approximate solutions for Z with no spatial variation, or variation in a segmented target. Section VII, making different approximations, discusses a method of using this method for high Z with spatial variation. Possibly it has some application to the plasma formed on the hohlraum walls. Section VIII discusses the method where the energetic particles are not from instability, but from the energetic tail of a Maxwellian distribution. Section IX gives some digressions, and Section X is the Conclusion.

## II. Two Relevant Instabilities

Two parametric instabilities which involve the electrons, and which have the potential to generate many fast electrons are the $2\omega_p$ instability, and the absolute stimulated Raman side scatter instabilities. Experiments have been performed on both the OMEGA laser at Rochester, and on the NIF laser at Livermore.

The threshold laser Irradiance for the former is given by

$$I_{th} = \frac{2x10^{16} T_e(keV)}{L_\mu \lambda_\mu} \frac{W}{cm^2} \qquad (3)$$

where $T_e$ is the electron temperature, $L_m$ is the electron density gradient scale length at the quarter critical density in microns, and $\lambda_\mu$ is the laser wavelength in microns.

The threshold for the latter is given by

$$v_{os}^2/c^2 = (k_o L)^{-4/3} (\omega_p/\omega_o)^{2/3} \qquad (4)$$

In terms of the laser irradiance and wavelength, the electron oscillating velocity divided by the speed of light is

$$v_{os}/c = 9x10^{-3}\lambda_{\mu}I_{14}^{1/2} \qquad (5)$$

where $I_{14}$ is the irradiance in units of $10^{14}$W/cm$^2$ (40, 41). Since the $2\omega_p$ instability can take place at any position in the under dense plasma with density less quarter critical, this depends on the plasma frequency $\omega_p$ and the laser frequency $\omega_o$. Generally, the instability occurs at about the quarter critical density.

Experiments on these instabilities were done at both URLLE with their OMEGA laser and at LLNL with NIF. Generally, OMEGA experiments, with their lower electron temperatures and shorter scale lengths excite only the $2\omega_p$ instability. NIF, on the other hand with its higher temperature and longer scale length excites the stimulated, absolute Raman side scatter instability. In all the experiments, the laser pulse is brought up from an irradiance of zero to some constant value which persists throughout the pulse. Their data has been presented as a measurement of this or that as a function of the steady irradiance. The parameters they measured were hot electron temperature and total energy of the energetic electrons as a fraction of total laser energy.

One iconic measurement of energetic electrons was performed at URLLE by a group led by Baruch Yaakobi. (42) It was with a planar target which was subject the $2\omega_p$ instability. The data points in Figure (5) show these measurements.

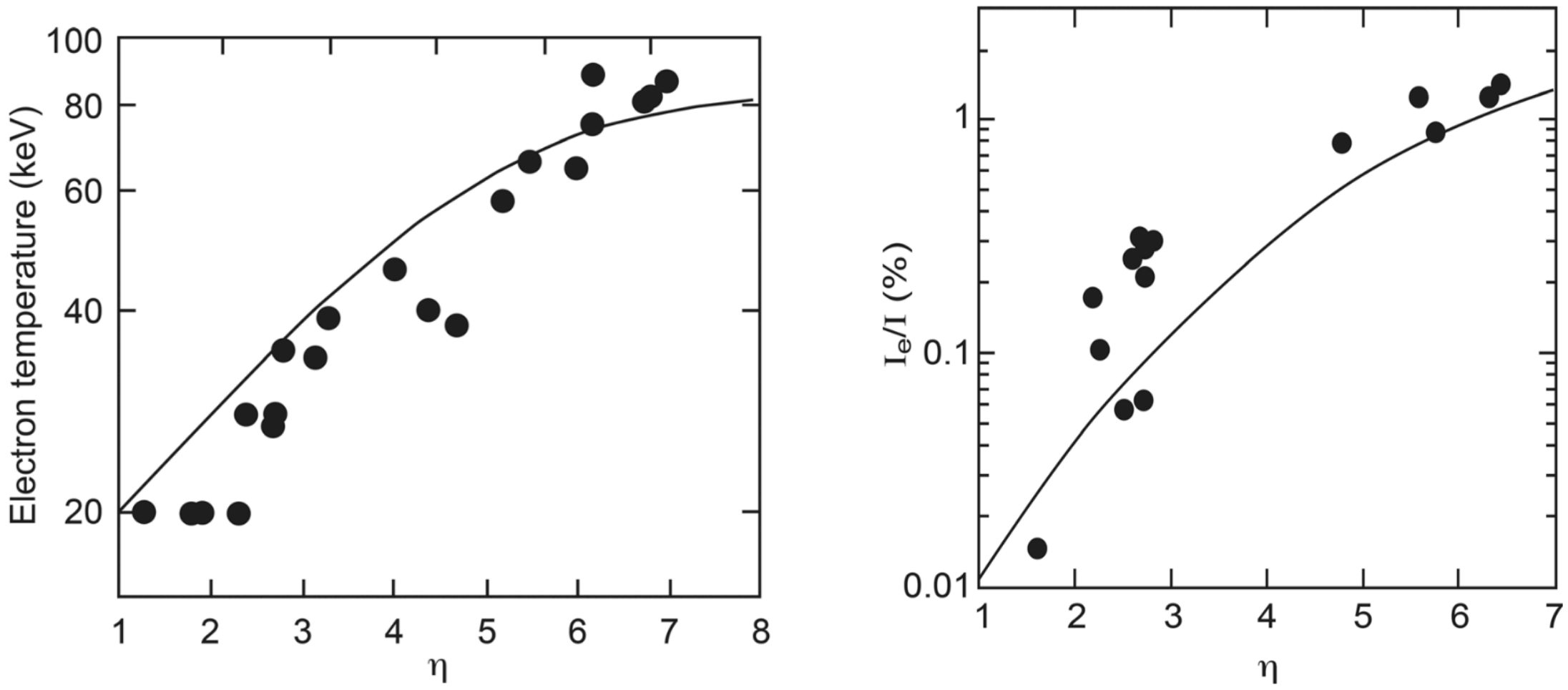

Figure (5). Data points from the URLLE experiments on the energetic electrons produces by the $2\omega_p$ instability. The horizontal axis label η, is the ratio of Irradiance to threshold irradiance. The left curve measures hot electron temperature, the right, ratio of electron energy to laser energy. The curves are empirical fits to the data.

The curves are:

For the temperature:

$$T_e(eV) = 2\times10^4\eta^{2/3} \quad 1 < \eta < 7. \tag{6}$$

And for the energy ratio:

$$\frac{I_e}{I} = 10^{-4}\eta^{2.5} \quad 1 < \eta < 7. \tag{7}$$

For the NIF experiment (27), here is the analogous data:

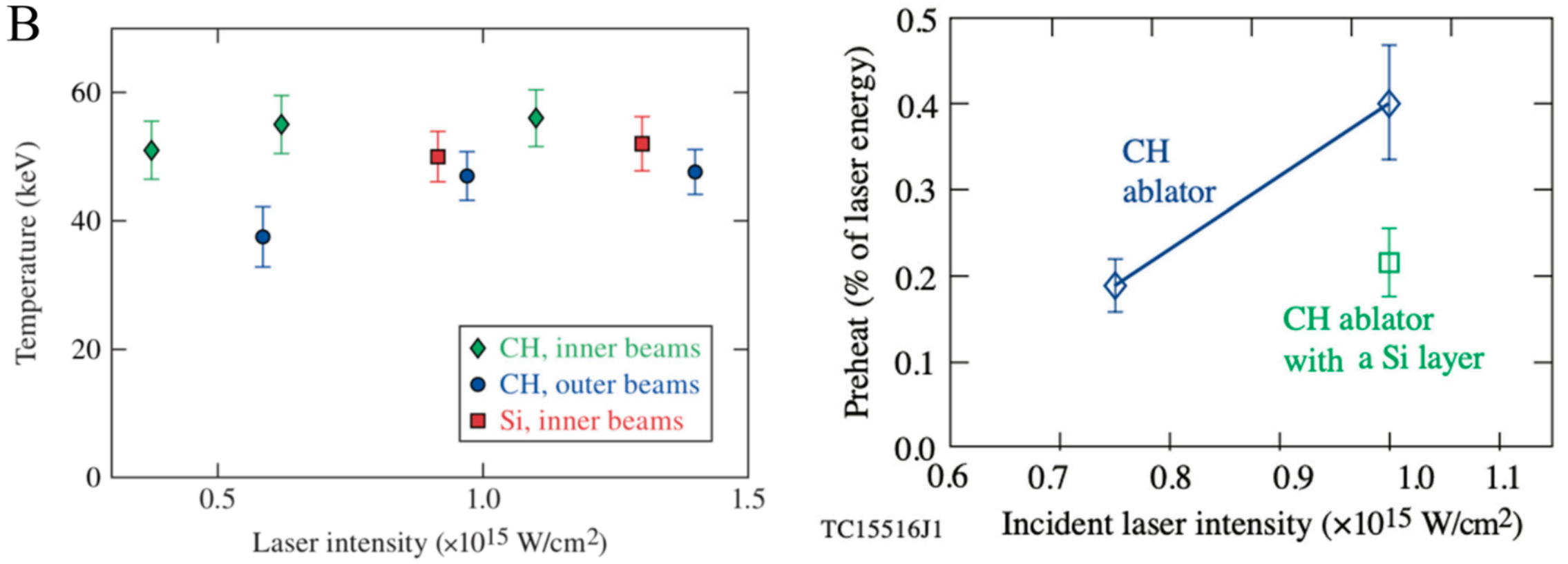

Figure (6) Data analogous to Figure (5). The URLLE/NIF group explored reduction of the instability by using different material in the ablating plasma. As a function of irradiance, the energetic electron temperature was relatively constant at ~ 50keV. The irradiance varied between ~ 0.5-1.5X$10^{15}$ W/cm$^2$.

One interesting thing about this data is that for the 2$\omega_p$ instability, the temperature rises with irradiance, whereas for the stimulated Raman, the temperature is relatively constant as the irradiance varies. Possibly the reason is that for the 2$\omega_p$ instability, only one of the wave numbers (that of the incident wave) is specified; there are large number of possibilities for the wave numbers of the stimulated electrostatic waves. Possibly, the size of region of the unstable wave spectrum in k space increases as the irradiance does. However, for the Raman case, the wave number of the two electromagnetic waves are basically specified, leaving little room for variation of the phase velocity of the stimulated electrostatic wave. Thus, one would expect that the energetic electron temperatures for the two instabilities might be quite different, with the Raman case having a temperature less dependent on irradiance.

Let us note that the data in Figures (5 and 6) are for steady state irradiance. However, in a laser fusion implosion, the laser irradiance varies very rapidly in time. The data above gives little hint of how to handle this in a rad hydro code like HYDRA of FAST. At this point the only reasonable possibility is to assume that in the rad hydro code, the instantaneous ratio of energetic electron energy flux to laser irradiance is equal to the ratios shown in Figures (5 and 6).

The next issue is the angular distribution of the energetic electrons. It is an issue where this author believes that the information gathered so far does not necessarily give the proper insight as to what is going on. One iconic experiment is a continuation of the experiment already sited (43).

In a different target configuration, there are several concentric spheres. The outer sphere is a thin shell which is irradiated pretty much the way the flat plane was in generating Fig. (5). The sphere is filled with nitrogen, and inside there is another sphere of high Z material, usually molybdenum. The hard X-rays generated depend on radius of the inner sphere, i.e. how much of the of the X-ray flux the sphere intersects. The experiment configuration, taken from (42) is shown in the left-hand image of Figure (7).

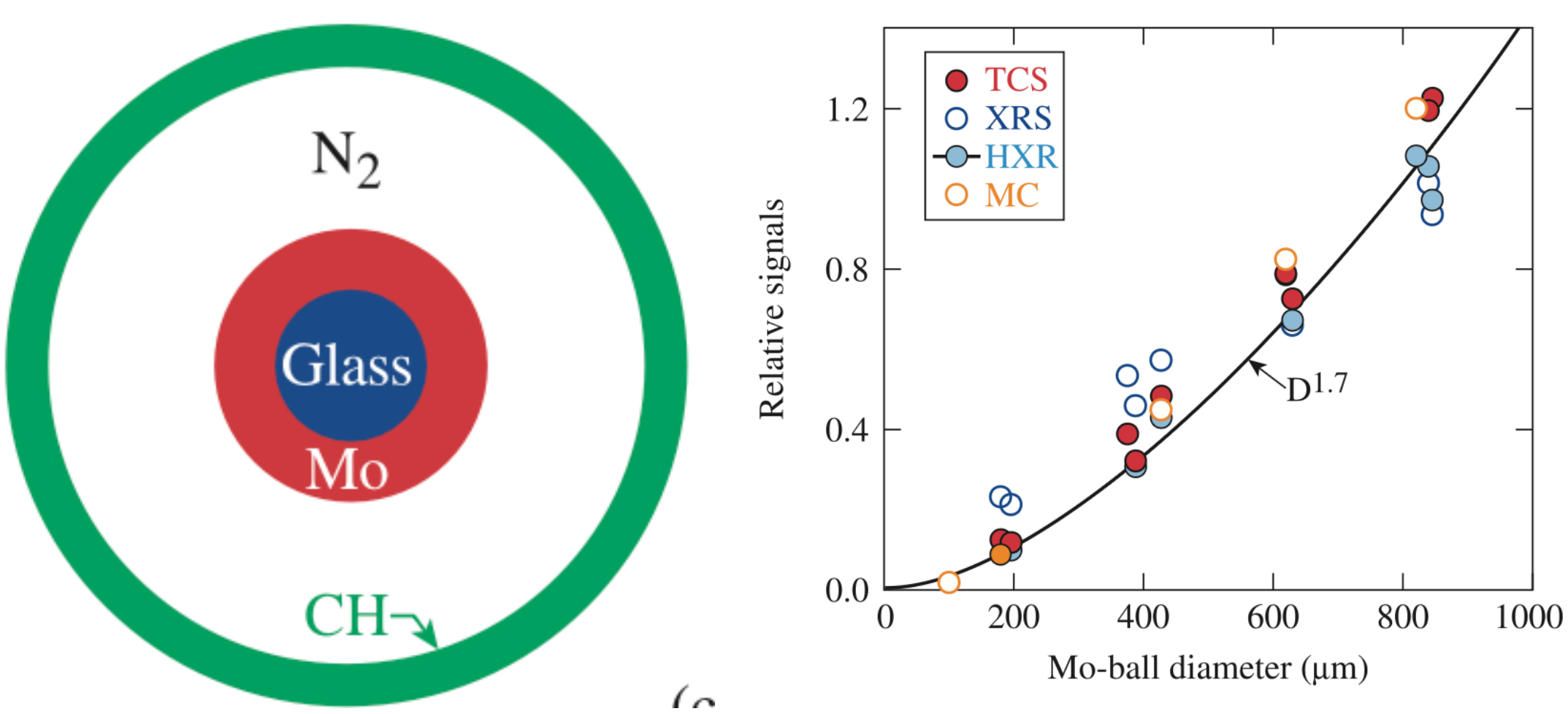

Figure (7) Left: The configuration of the experiment, (43); Right: the total X-ray energy as a function of radius of the Molybdenum ball.

Notice from the center graph that the total hard X-rays increase roughly as the ball radius squared, indicating that the electrons come out in a wide angle.

There are also results from PIC simulations (44). This work took very different parameters for both the laser and plasma. However their simulation was subject principally to the Raman side scatter instability, and they certainly observed it. They observed a tail on the distribution function having a temperature of about 50 keV, just like that seen in the NIF experiment. Furthermore, the angular spread of the energetic particles was confined to a cone with about a 45-degree vertex angle. These results are displayed in Fig (8). The plot of the distribution with energetic tail, was taken from (44). The phase space plot did not appear in the publication. However, Dr Xiao was kind enough to send me his plot and allow me to use it in my work. For this, I am certainly grateful.

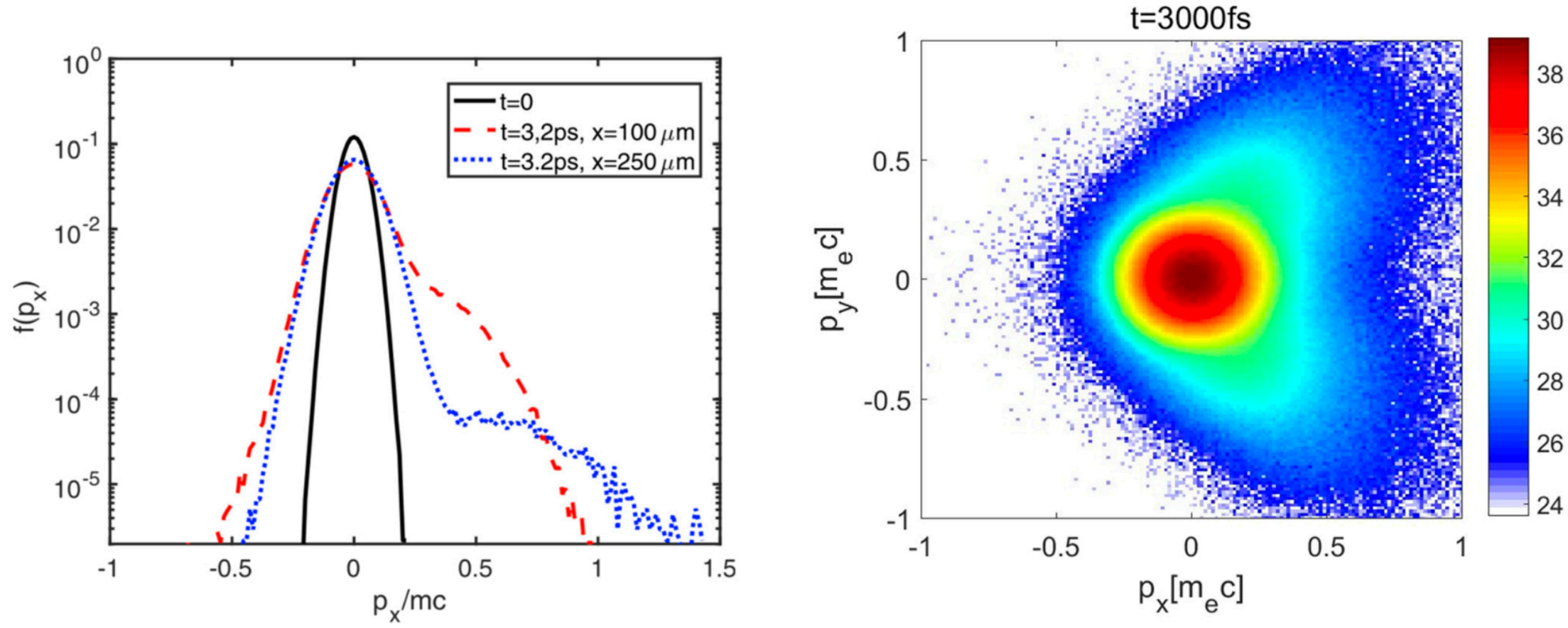

Figure (8) Left: A plot of the initial distribution function in black, a 1 keV plasma. The red and blue curves are distribution functions at 2 positions in the plasma after 3.2 psec. The blue curve is near the higher density edge of the plasma. Clearly it is an energetic distribution with a temperature of around 50keV and with a density of 3 or 4 orders of magnitude below the main plasma density, consistent with the NIF results shown in Fig (6). Right: A phase space plot of the energetic electron distribution. They are confined to a cone with about a 45 degree vertex angle, basically consistent with the results of Ref (43) (but for a different instability) and Figure (7)

However, despite the impressive nature of the experiment, and the simulation of the instability with a PIC code, this author feels that the impressions left in the literature are not especially helpful for the case of a direct drive laser fusion target, i.e. just the configuration which this manuscript attempts to enlighten. In the real world of direct drive laser fusion, the plasma ends at neither the last cell of the simulation, nor the edge of a thin spherical shell. In the actual case, the quarter critical density (for instance) abuts a very inhomogeneous plasma whose density just beyond the quarter critical density rises very rapidly, finally reaches a maximum at much higher density, and then reduces until it is near the center of the target.

We take here an example from (27).   Figure (9) shows a plot of density versus radius for a rather complex, striated plasma taken from (27), and reproduced in different for the different Z plasma below  in Fig. (9).

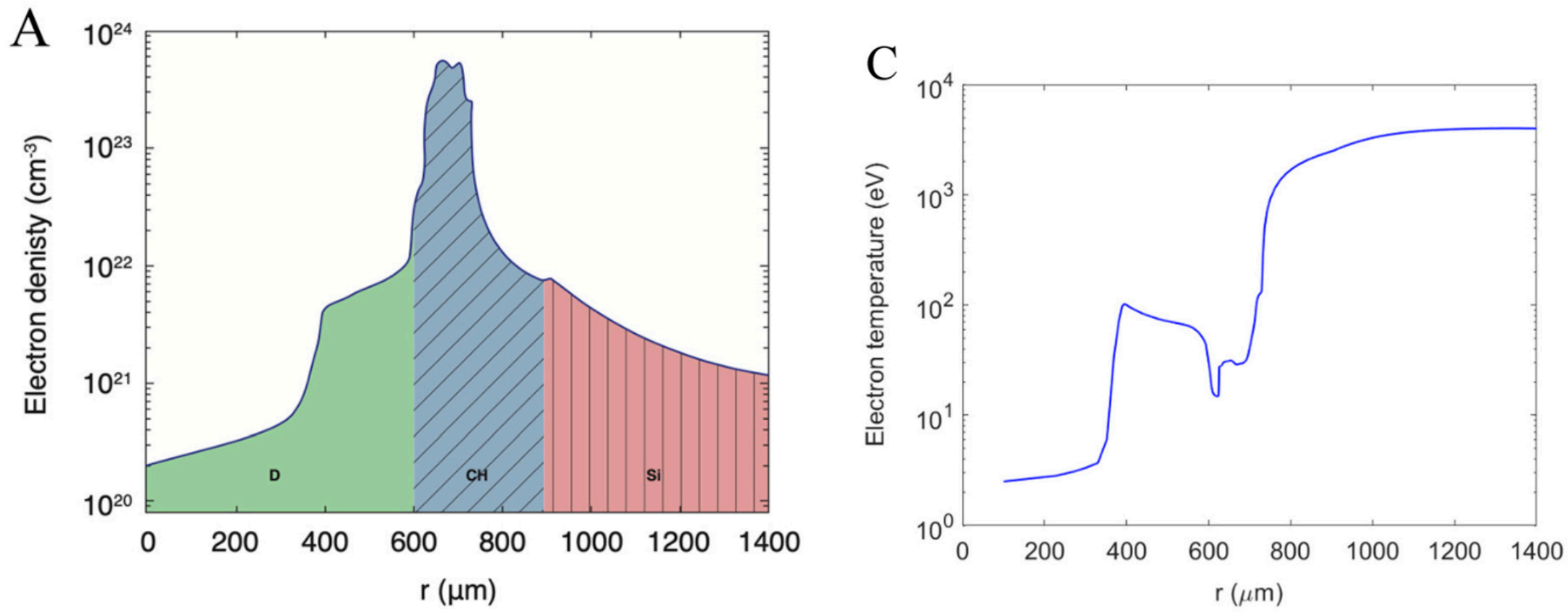

Figure (9):  A plot of the density (A) and Temperature (C) of the plasma as calculated in (27) at a particular time of their simulation.  The plasma has 3 regions, an outer region of silicon, (i.e. Z=14) an inner region of CH (i.e. plastic, Z=5.3), and the target region of deuterium (Z=1).  The quarter critical density is near the right-hand edge.

Once the energetic electron (say 50 keV) gets beyond the quarter critical density and into the denser plasma, the electron hardly moves along straight-line orbits.  However often it is portrayed in the region as doing just that.  For instance, look at Figure (10):

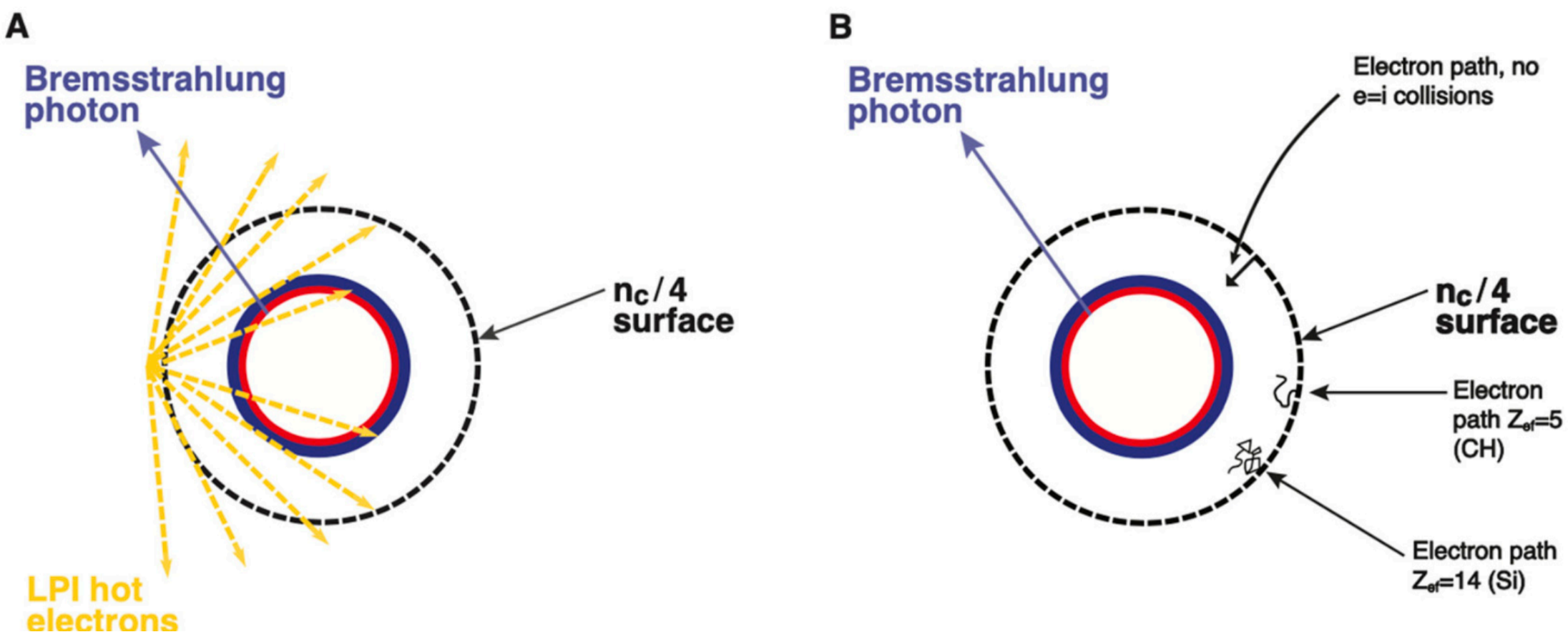

Figure (10).  Left hand image:  The paths of energetic electrons as envisioned in Ref (27), the source of this image.  Right hand image:  A more accurate picture the path of a 50-100 keV electron once it passes just inside the quarter critical density.

The arrow in the upper right hand side of the right hand image is the path of a 50 keV electron, moving directly toward the center, and moving an electron-electron mean free path at for a density profile like that given in Fig (9) (actually as we have seen, the actual path it travels before it reaches zero velocity is much shorter). However, even for Z=1, this is not an accurate representation. The electron scatters in angle from both the ions and electrons. Hence before the electron travels a mean free path, it will have made several sharp turns in its orbit. The case for a Z=5 (plastic) and Z=14 (silicon) plasma is shown schematically in Fig (10B) as well.

Clearly envisioning energetic electrons as moving along straight-line orbits through the center of a laser fusion target does not give an accurate picture of what is really happening, at least not for those electrons that preheat the fuel. The question is how does one describe the energetic electron distribution just inside the quarter critical density? Clearly these electrons are attempting to become a spherically symmetric distribution due to both electron and ion collisions. The collisions approximately conserve electron energy, more accurately for higher Z ions, but even approximately so for a Z=1 plasma. However, there is a constraint. Since the collisions are mostly with the heavier ions, they do not lose any energy. Therefore, the energy flux of these energetic electrons just inside the quarter critical surface must be the same as it was just outside; otherwise, there would be enormous heating at the quarter critical surface. Hence the calculation to do is to minimize the anisotropy, but subject to the constraint that the energy flux is as given is as given like in Fig (8) at each energy. (This implies a higher density of these energetic electrons than what the simulation produced). This seems like a normal variational calculus problem. However instead of performing this calculation, the author simply used a drifted Maxwellian, which is a certainly reasonable approximation. As this work is formulated, the only distribution function needed, when expanding f in a Legendre series is the first term, the term proportional to cos θ, the cosine of the angle between the particle velocity and the spatial gradient. For that we take

$$f_1(v,\theta) = Av\cos\theta \exp - v^2/{v_{th}}^2 \quad (8)$$

Here A is the coefficient that gives the proper energy flux for the energetic electrons as calculated in the simulation (44) (for instance).

One way to solve the deposition problem is to use a Monte Carlo code and follow the jagged path of many, many electrons starting at the quarter critical surface, and ending when the electron had lost all its energy. Clearly, to complete the task, many, many thousands of electrons would be needed both to fill up the distribution function, and to calculate every type of orbit in the spherical plasma. That is an approach that Fig (10A) hints at. But can this be economically done at every time step of rad hydro code?

However, in the next few sections, we will find that one can short circuit this approach by approximately solving the Fokker Planck equation for the energetic electrons. Getting such an approximate solution, using one of two different approaches, is simpler than one might think.

## III Fokker Planck (and other) approaches

There are many simplifications one can make on the actual Fokker Planck equation, which are relevant to the problem at hand, namely the preheat of a direct drive laser fusion target. These have been described elsewhere (23-26) . There are two types of problem for which the Fokker Planck treatment can be optimal. First, the energetic electrons could be produced by the one of the laser plasma instabilities, like those just summarized in the previous section. Second, they could be the tail of a Maxwellian distribution, that is those thermal electrons with high enough energy that they mostly decouple from the body. Many of the approximations and analyses we make here are common to both situations. Others are applicable to one or the other.

We begin with the case of instability generated energetic electrons and then continue in a later section by discussing the case of the mostly decoupled tail of the distribution function. There are two approximations we make, which, and almost certainly correct. First, we assume that the number of energetic electrons is very small, as compared to the total plasma. Immediately this greatly simplifies the theory as compared to nearly every other one which deals with the entire electron distribution function.

Figures (5 and 6) in the last section shows that experiments up to now have shown that the energy of these particles maximizes (at least in previous recent experiments) at an energy of about 1% of the laser energy, or as we more accurately put it for our configuration, the energy flux of the energetic particles, so far maximizes at about 1% of the laser irradiance. (However, this 1% or less, might have a profound effect on the preheat of the target.) Hence, these particles interact with the background plasma and hardly interact with one another. While the overall Fokker Planck formulation is strongly nonlinear, the Fokker Planck calculation of the interaction of these particles with the plasma is a *linear* problem; double the number of energetic electrons, and the preheat of the target doubles.

Secondly, as shown in (23), the Fokker Planck interaction terms are characterized by two of what are called Rosenbluth potentials. One is labeled F, which is a vector and is basically a friction force, the other, labeled D is a tensor and it has elements of both friction and diffusion. However, if the particle is energetic compared to the background, these terms simplify very considerably. It turns out, as shown in (23) and other places, that the collision of the energetic electrons with the ions contains only angular scattering with a coefficient proportional to the effective Z of the plasma (i.e. the electron energy is conserved). The interaction of the energetic electron with the background electrons has two components. First the background electrons give rise to angular (i.e. energy conserving) scattering equal to that of the ions if Z=1. Second the collision gives rise to a friction force which slows the electron down. As we have seen in the Introduction, the electron cannot even travel one electron-electron collisional mean free path (mfp) before it stops. Finally, the Diffusion tensor describes the diffusion in energy of the energetic electrons as they collide with background electrons. However, for energetic electrons, this process is smaller than the other processes just described by about a factor of the background electron temperature divided by the energetic electron energy. This is fine for the instability generated electrons but is certainly less accurate for the energetic electrons arising from the tail of the Maxwellian, at least near the position where these electrons are born. (For much of their life, they traverse a much cooler plasma.) We neglect this electron energy diffusion in this work.

To reiterate, in a purely Krook calculation, after the electron travels one mfp, it has a probability of 1/e of remaining and not joining the thermal plasma. For this reason, earlier Krook calculations, by NRL (15), URLLE (16) and probably other organizations, have shown no gain in laser implosion simulations. As far as this author knows, none of these calculations which show failure to ignite have been published.

Different organizations realized that friction had to be accounted for, as discussed in the Introduction. Thus, they added terms to the Krook model which they hoped would accomplish this. This author thinks it is better to take a theory which is correct and work with it, rather than take one that is basically incorrect and attempt to patch it up. To be honest, the NRL group was probably the last to realize the importance of friction. We learned our lesson the hard way by performing several rad hydro simulations with a Krook model for the energetic electrons. We never found satisfactory implosions with acceptable gains (17). Apparently URLLE never did either (18). However, at this point, rather than attempting to find a bandage, we dropped the Krook model and moved onto a Fokker Planck model. We believe this model is basically correct and that where the approximations we have made are not satisfactory, they can be corrected with further effort and study.

To get back to our analysis, we also simplified the Fokker Planck model by looking at steady state solutions for the electron energy flux, i.e. the theory is nonlocal in space, but steady state in time. The plasma density, energetic electron energy, and electron collision times seem to easily support this approximation (24,26). However, doing the calculation only in space means that we must specify spatial boundary conditions. At the center of the sphere, (or at the analogous end point in a planar approximation to the sphere) there obviously cannot be any energy flux at any particle velocity. For the instability case, the other point is at the quarter critical density and the energetic electron distribution function is given by Eq (8) with A and the energetic electron temperature given in the last section.

Finally for the case of the instability, we neglect the effect of the electric field. For the simple spherical or planar geometry, the electric field simply the gradient of a potential. The potential across the plasma is generally of order of the electron temperature a few keV. However, the instability generated energetic electrons have a temperature of ~ 50 keV, and many electrons have energy in the hundreds of keV. It is interesting to note that with our simplified FP formulation, there are no complicated numerical issues. In fact, we have examined the numerical stability of the rad hydro energy equation with such an additional heating and found it is nearly always stable (11). The added term to the rad hydro equation is a simple heating term (actually it is the divergence of an additional heat flux). This contrasts with the more complex multi-dimensional calculations, containing many additional effects, i.e. generated magnetic fields, this or that instability…For instance in one case, the code had to occasionally use an artificially reduced electric field (22):

In some circumstances with strong temperature gradients, (that is the electric field across a grid cell is greater than the collisional slowing down) , meaning that it is impossible to meet the CFL condition described in Eq. **(21)**. In these circumstances, the code will always crash, regardless of whether the diffusive term is used or the number of energy groups used. Here, an artificial reduction in the electric field experienced by higher energy groups can be used to stabilize the M1 code.

There are other places in Ref (22) where they had to add specific limitations to keep their code from crashing.

For our formulation, the electric field given strictly by the gradient of a potential, and our method of solving the for the additional energetic electron energy flux, the electric field may be large enough to violate the condition above (22) between two collisions, but it makes no difference. Look at the orbits as visualized by Figure (10). No matter how much the electron gains energy in one jag of its orbit, in the next jag, that energy is given back. The effect of the electric field at from the initial to final point of the orbit, as shown in Fig (10), is given only by the potential drop between these two initial and final point, i.e a keV or two out of 50 or more keV particle energy.

For the case of the nonthermal tail, the outer boundary condition is more difficult and subjective. Also, one can still neglect the effect of the electric field on the energetic tail, by the same logic as above, but validity of this approximation becomes much more questionable. We will discuss this in a later section.

While the formulation here is one dimensional, planar or spherically symmetric, there is always the potential of extending the theory to 2 or 3 dimensions. However, the author does not believe that this is a high priority item, at least at present for the case at hand. If the multi-dimensional rad hydro code predicts acceptable deviations from spherical symmetry for ignition, one can always calculate the spherical average of each quantity and then calculate the spherically symmetric average of the effect of the energetic electrons. If the multi-dimensional rad hydro code shows large deviations from spherical symmetry, it is unlikely that the configuration will be acceptable for either laser fusion or additional study; and if not, who cares what the effect of the energetic electrons would have been.

Now to get to the nitty gritty of the Fokker Planck analysis. The Fokker Planck equation, with all the simplifications we have just mentioned becomes

$$\frac{\partial f}{\partial t} + V\cos\theta\frac{\partial f}{\partial r} - \frac{V\sin\theta}{r}\frac{\partial f}{\partial \theta} = \frac{1}{V^2}\frac{\partial}{\partial V}V^3\nu_e(V)f + \frac{\nu_e}{2}(1+Z)\left[\frac{1}{\sin\theta}\frac{\partial}{\partial\theta}\sin\theta\frac{\partial}{\partial\theta} + \frac{1}{\sin^2\theta}\frac{\partial^2}{\partial^2\phi}\right]f \tag{9}$$

It is appropriate for either a planar situation (the r variable would be denoted by x or z) or a spherical configuration. The variable θ is the angle between the velocity vector and the vector in the direction of the gradient, in the planar or spherical configuration. The third term on the LHS accounts for the fact that the vector between the velocity and spatial gradient has a spatial dependence in the spherical case. The variable $\nu_e$ is the electron collision frequency given by

$$\nu_e = \frac{4\pi n e^4}{m^2V^3}\Lambda \text{ in cgs or } \quad \nu_e = 3.9\times 10^{-6}\frac{n}{\Xi^{3/2}}\Lambda.$$

Here, Λ is the Coulomb logarithm, and Ξ is the electron energy in electron Volts. The next step is to expand the distribution function in a series of Legendre polynomials keeping only the first two terms ($P_o(\theta)=1$, $P_1(\theta)=\cos\theta$) (we discuss additional terms in the expansion in a later section):

$$\frac{\partial f_0}{\partial t} + \frac{V}{3}\frac{\partial f_1}{\partial r} + \frac{2V}{3r}f_1 = \frac{1}{V^2}\frac{\partial}{\partial V}V^3\nu_3(x,V)f_0,$$
$$\frac{\partial f_1}{\partial t} + V\frac{\partial f_0}{\partial r} = \frac{1}{V^2}\frac{\partial}{\partial V}V^3\nu(x,V)f_1 - \nu_e(x,V)(1+Z)f_1. \tag{10}$$

Various terms in Eq (10) are stuck out by red solid lines, or green dashed lines. The former are terms that are neglected in the steady state approximation. The latter are terms struck are what is called the multi group diffusion approximation. This is generally used to calculate neutron transport (19), but it has also been used to calculate energetic electron transport in laser produced plasmas (18). The author has always been suspicions of the multigroup diffusion approximation. After all, why eliminate dynamical friction in the equation for $f_1$, but not in the equation for $f_o$, and why eliminate the time derivative in the equation for $f_1$, but not in the equation for $f_o$? Here we consider only the steady state approximation.

To simplify things, we simplify the definition of variables:

$$dz = \sqrt{3}\frac{2.7 \times 10^{-13} n(r)\Lambda}{\mathrm{T}_e^2} dr \equiv k(r, T_e)\mathrm{dr} \quad \text{and} \quad w = \varsigma^2 \equiv \left[\frac{\Xi}{T_e}\right]^2. \tag{11}$$

Here, $T_e$ is the temperature of the energetic electron distribution in electron Volts and z is defined as zero at the spherical center (or its equivalent position in the planar approximation to the spherical configuration). Notice that there are no dependences on Z in either of these variables. All Z's are explicitly noted in the equations. A change in z of unity means a single electron-electron mean free path. Recall that the electron can only travel about a quarter of a mfp before it stops completely. As an example of the dependence of z on r, Figure (11) shows the calculation of z for the density distribution shown in Fig (9) (the spatial dependence of the temperature is also given in Fig (9)).

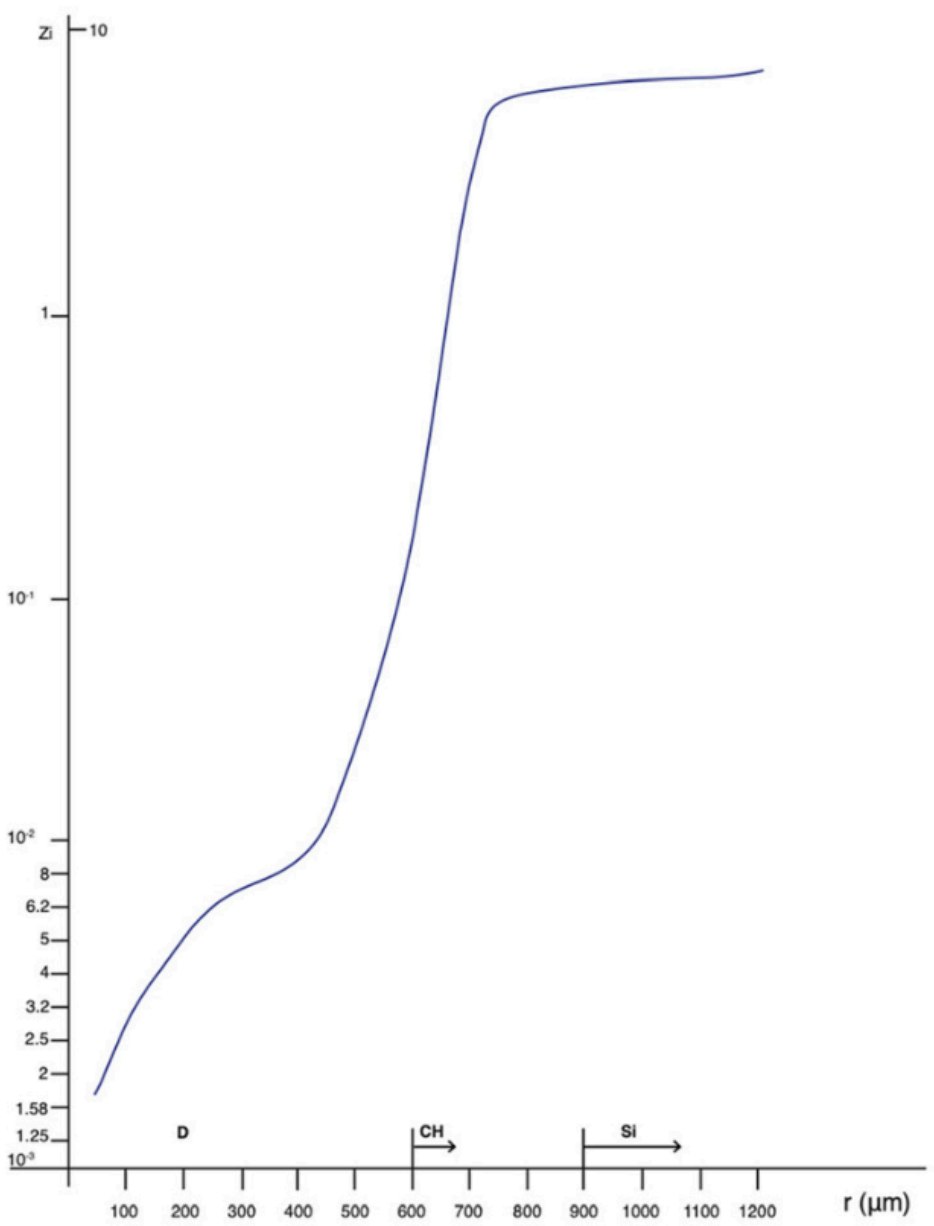

Figure (11): A plot of z versus r for the electron density given in Fig (9). The electron mfp varies tremendously in the plasma. The total size of the plasma is about 6 of the electron mean free paths at $T_e$ .

Notice at r=1200 microns, z is about 6, and it decreases to about 5 at about 800. This is approximately consistent with the sketch of orbits in shown in Fig. (10).

Then eliminating the equation for $f_o$, we find a single, second order partial differential equation for only $f_1$.

$$\frac{\partial^2 f_1}{\partial z^2} + 2\frac{\partial}{\partial z}\frac{f_1}{r(z)\left[\frac{dz}{dr}\right]} = \frac{\partial^2 f_1}{\partial w^2} - \frac{1+Z}{4w}\frac{\partial f_1}{\partial w} + \frac{1+Z}{4w^2}f_1. \tag{12}$$

This is the equation (and various approximations to it), which we will work with, in most of the rest of this manuscript. The next four sections develop methods of solving Eq (12). The next section makes some approximations to the equation and then find an exact solution. In following sections, we keep the equation as it is (well almost) and develop methods to find an approximate solution.

## IV. Solution of the Fokker Planck Equation by Characteristics

We notice that Eq. (12) looks a lot like the wave equation, but with a few extra terms; two on the right where the Z dependence comes in, and one on the left where the spherical nature comes in. For now, we neglect the spherical the spherical nature of the configuration, so the second term on the left disappears. In the next section we discuss this further.

We start by assuming the plasma is characterized by a single Z. Then we consider a segmented target like that shown in Fig. (9). The equation simplifies somewhat if we can rewrite the rhs without the first derivative. Redefining the dependent variable as g rather than as $f_1$, where

$$f_1(W,z) = W^n g(W,z) \tag{13}$$

When n=Z+1, we get the equation for g:

$$\frac{\partial^2 g}{\partial z^2} = \frac{\partial^2 g}{\partial w^2} - \frac{(1+Z)(9+Z)}{64w^2}g. \tag{14}$$

For a rather decent part of the parameter space, especially for lower Z and larger w, we find that we can simply neglect the last term (we will define this more precisely shortly). Then the equation simply becomes a wave equation for g;

$$\frac{\partial^2 g}{\partial z^2} = \frac{\partial^2 g}{\partial w^2} \qquad (15)$$

Now let us consider the boundary conditions. The relevant region of space $0 < z < z_c$, where $z_c$ is the quarter critical density, and for w between zero and infinity (i.e. positive values). We know (from Eq(8)) that at $z = z_c$,

$$g(w, z_c) = w^{-(Z+1)} w^{1/4} exp(-w^{1/2}) \equiv g(w) \qquad (16)$$

This is one boundary condition. Another is that g = 0 at z = 0 for all w (i.e. there can be no energy flux into the origin at any energy). Furthermore, we also know that g = zero for w equals infinity. The z axis, i.e. w=0 does not come into play as we will show shortly.

These are odd boundary conditions for a wave equation. It is unusual for a wave equation for g, to have the boundary condition specifying g on a *closed* surface. It is more usual for it to be specified on open surface, say $z=z_c$. However, one also would need to know the partial derivative with respect to z on this surface to determine a solution.

Nevertheless, as we will show, the physical conditions do give a unique and physically reasonable solution. We know that there are two characteristics from the wave equation emanating from the point on the boundary $(w_o, z_c)$. They are

$$w - w_o = z - z_c \quad \text{and} \quad w - w_o = -(z - z_c) \qquad (17)$$

These two characteristics are illustrated in Fig (12), emanating in two directions from $w_o$. Notice that the ones in black proceed downward as z decreases from $z=z_c$, while the ones in red proceed upward. Obviously, it is completely unphysical that there can be any solution on this upper characteristic, so the solution is fully determined by forbidding any solution on this characteristic. To relate it to the more normal boundary condition, it is not difficult to see that this new condition is completely equivalent of setting the derivative boundary condition as

$$\frac{\partial g}{\partial z} = \frac{\partial g}{\partial w} \quad \text{at } z=z_c \qquad (18)$$

(Correspondingly, if the only allowed solution were on the upper characteristic, the relevant normal boundary condition would be Eq. (18) but with the sign reversed)

Hence, we simply express this as saying that only one of the characteristics can be occupied, the one in black in Fig. (12).

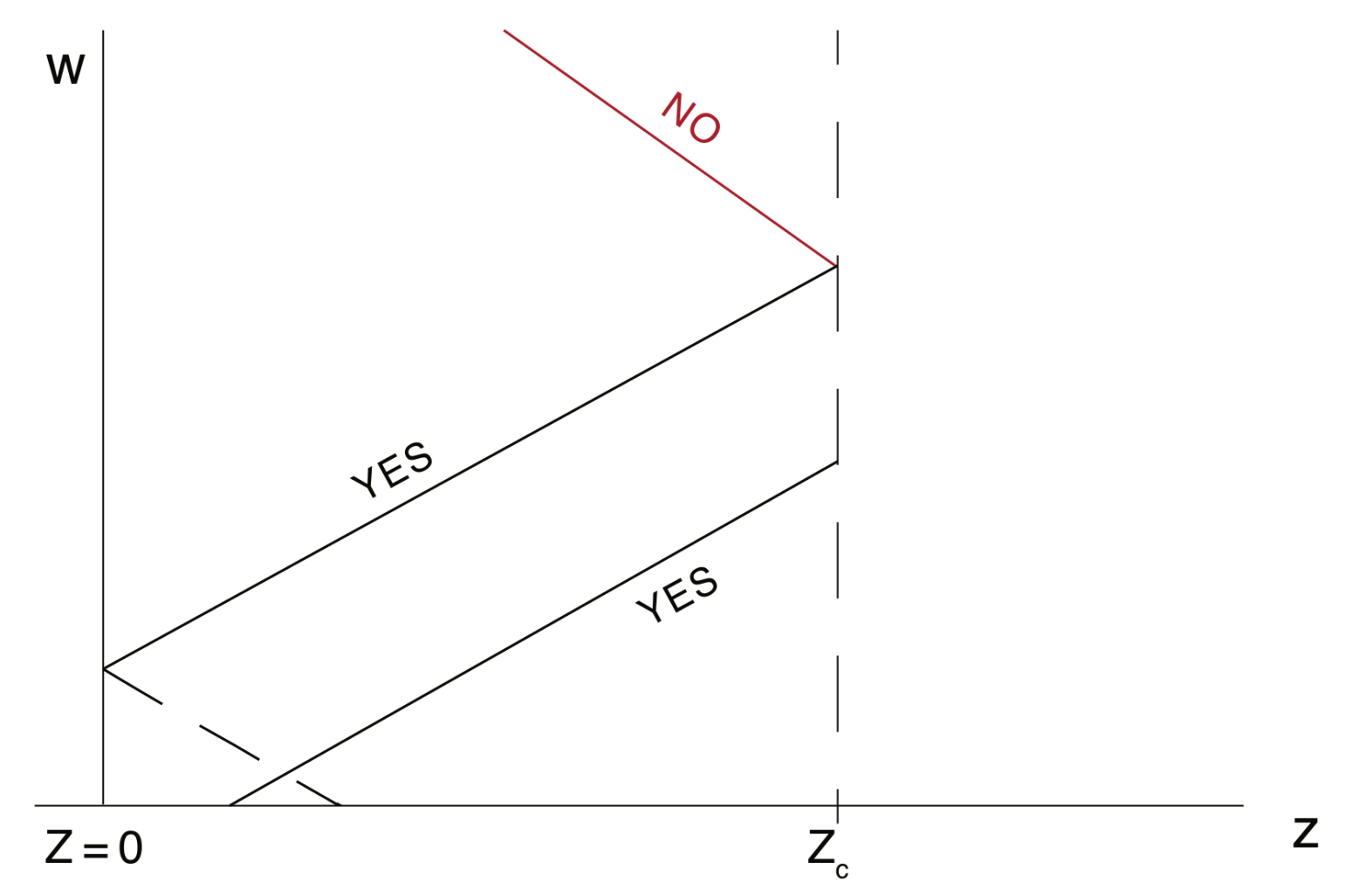

Figure (12) The region of interest in wz space for which there could be significant heating from the energetic electrons. Clearly there can be no solution for the red characteristics, as the energetic electron energy would increase from its value at $z=z_c$ as it traveled along this upper characteristic.

As the solution moves down the characteristic from its position at $z=z_c$, there are two possibilities. The lower characteristic in Fig (12) strikes the z axis, i.e. w=0. In this case, the calculation on this characteristic simply ends; the energetic electron has given up all its energy and can no longer be considered The other possibility is that the characteristic strikes the w axis, i.e. z=0. This would violate the boundary condition at z=0. However, this can be remedied by adding a 'reflected wave', originating at the intersection which 'propagates' back toward the quarter critical surface, but with negative coefficient. In other words, the solution to the 'wave equation' can be expressed analytically as

$$g(w,z) = g(w + z_o - z) - g(w + z_o + z) \qquad (19)$$

This is a valid solution as long as w>0. If $w_o$ is small enough that characteristic strikes the z axis (i.e. w=0), the second term of Eq. (19) does not appear. The second term only appears for characteristics that strike the w axis (see Fig 12). If Eq. (19) give a w<0, and/or z=0, we simply ignore it. In other words, as expressed in terms of $f_1(w)$, the value of $f_l$ at $z=z_c$,

$$f_1(w,z) = \left[\frac{w}{w+z_c-z}\right]^{Z+1} f(w+z_c-z) - \left[\frac{w}{w+z_c+z}\right]^{Z+1} f(w+z_c+z) \quad (20)$$

for w >0 and $0 < z < z_c$. Again, the second term appears only for characteristics that strike the w axis.

Once we have the $f_1$, it is simple enough to get the energy flux as a function of z:

$$q_{ee}(z) = \int dw w^{1/2} f_1(w,z) \quad (21)$$

Recall that A in Eq. (8) is chosen so that the energetic electron energy flux at the quarter critical density has the proper value.

Now let us briefly examine validity of approximating Eq (14) by the wave equation. Simply taking the second derivative of the exponential part of Eq (20), i.e. the dominant term at high energy, and comparing it to the second term on the right-hand side of Eq. (14), we find that the wave equation ought to be ok if

$$\frac{(1+Z)(9+Z)}{16w} < 1 \quad (22)$$

Clearly the wave equation works best for low Z and high w. Expressed as a function of w, the energy flux goes as $w^{3/4}$exp-$w^{1/2}$dw. This has a rather broad maximum, with significant energy flux extending from about w~2 to w~15. While the wave equation might not be the best way to get an approximate solution it does have one undeniable advantage. It is extremely simple. Using Eqs. (20 and 21), and taking the divergence of the energy flux, one has a simple expression for the energetic electron heating from quarter critical surface to the center of the target. It is a simple heating term which should not add any requirements to the rad hydro code, and it is unlikely that it would do anything to crash the code either. Once one has the necessary information on the effect of the instability, as described in for instance Section (II), it would be extremely easy to put this formula for energetic electron energy flux into the rad hydro code. This way, one could get what is probably at least a reasonable first approximation, at each rad hydro time step, to the effect of the energetic electrons on the target gain. It is almost certainly much easier to implement than calculating thousands of orbits at each time step with a Monte Carlo code.

In later sections we will see how one can find better solutions for Eq (14), including both the second term on the rhs of Eq (12), and the spherical effects.

Now we consider a segmented target, like that shown in Fig (9). Solving the wave equation for g in each segment, we need only the relation between the z at the boundary of between the region, i.e. at $z=z_{12}$ in Fig (13)

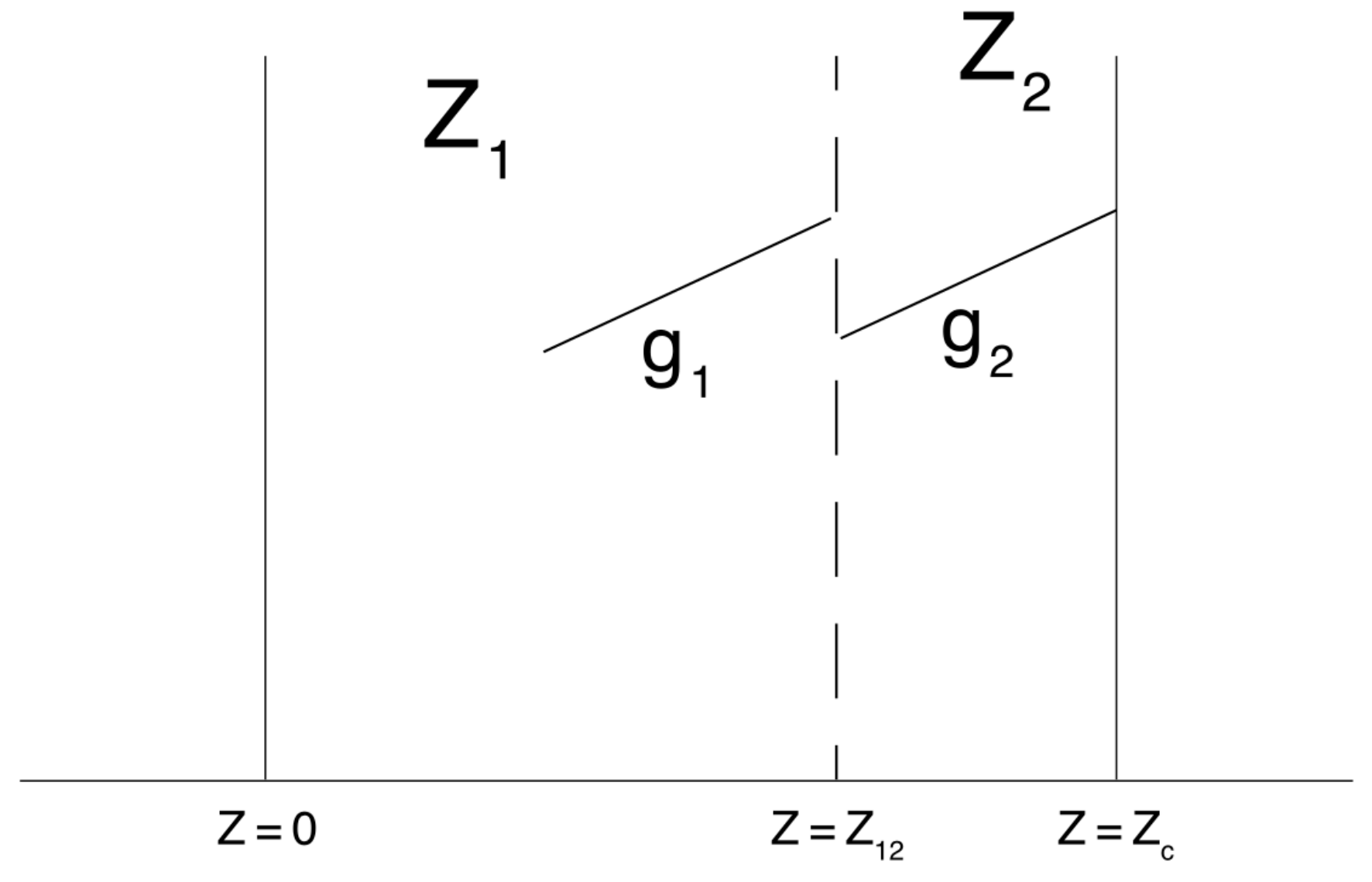

Figure (13) A figure of the characteristics for a segmented target. The boundary between the regions of $Z_1$ and $Z_2$ is denoted as $z_{12}$. While g(w,z) is not continuous across this boundary, f(w,z) is.

Since the energy flux across this boundary must be continuous, the discontinuity in g at the boundary is given by

$$w^{Z_1+1}g_1(w,z_{12}) \text{ at } z_{12}-\varepsilon = w^{Z_2+1}g_2(w,z_{12}) \text{ at } z_{12}+\varepsilon \qquad (23)$$

and the solution propagates downward along the characteristic in the next region.

To summarize, using the wave equation approximation to get an approximate solution to Eq.(12), we start at $z = z_c$ with the known function of $f_1(w)$ and from that get g(w) at $z = z_c$. We then propagate the solution for g down the lower characteristic, as shown in Figs (12 and 13), until the characteristic strikes either w=0 or z=0. For the former, the calculation along this characteristic simply stops; the energetic electron has lost all its energy. If the characteristic first strikes the z=0, one must add a 'reflected wave' with negative coefficient to satisfy the boundary condition, g(w,0) =0 at z=0. If the target is segmented, as shown in Figs (9 and 13), one must account for the different Z's occurring in the different regions, as shown in Fig (13).

V. An improvement to the characteristic approach

We have seen that the equation for $f_1$ can be rewritten as an equation for g, where g is defined by Eq. (13) for the planar case. Let us now consider the spherical case. In this case

$$\frac{\partial^2 g}{\partial z^2} + 2\frac{\partial}{\partial z}\left(\frac{g}{rz'}\right) = \frac{\partial^2 g}{\partial w^2} - \frac{(1+Z)(9+Z}{64w^2}g \qquad (24)$$

Here z' = dz/dr. To continue, let us define

$$g(w,z) = H(z)h(w,z) \qquad (25)$$

and define H in such a way as to make the first derivative term on the left hand side disappear. This will leave Eq. (24) as the wave equation, but with a correction term on the right-hand side which we will define as $P(w,z) = [\chi(w) + \Lambda(z)]h(w,z)$. Note that P is proportional to h(w,z). Clearly $\chi(w)$ is the added fraction on the rhs of Eq (14). We make one additional assumption, namely that H(z) is a slowly varying function so that second derivatives of it can be neglected. Then it is not difficult to show that

$dH/dz = =(rz')^{-1}H$ and $$H(z) = exp - \oint_{z_c}^{z} dz'\left(\frac{1}{rz'}\right) \qquad (26)$$

Then one can show that

$$\Lambda(z) = \frac{2}{(rz')^2}\left\{1 + \frac{d(rz')}{dz}\right\} \qquad (27)$$

So that the modified wave equation becomes

$$\frac{\partial^2 h}{\partial z^2} = \frac{\partial^2 h}{\partial w^2} + \epsilon P(w,z) \qquad (28)$$

Here ε is not necessary, small, but is introduced here as a bookkeeping device to be able to follow the perturbed quantities. To see the actual modified wave equation, simply set ε = 1.

Assuming that ε is small, we can approximate the rhs correct to order e by simply using the zero order solution for h (i.e. the solution to the wave equation, i.e. the solution h(w,z) is constant, along the downward characteristics that propagate to the left from $z=z_c$ ). Taking this solution, as an approximation to the rhs, we can solve for the modified wave equation, correct to order ε, by using a Green's function.

For the wave equation, we usually need two Green's functions, one for h(w,z) and one for its partial derivative with respect to z, since each usually must be specified to get a solution.

However, as we saw in section (IV), an alternative, and completely equivalent approach is to specify that there is no solution on one of the characteristics. In fact, certifying this is totally equivalent to specifying the partial derivative as written in Eq. (18). Hence for our case, we need only to specify a single Green's function, for h(w,z) itself, and simply eliminate the solution that it specifies on the unphysical characteristic.

Hence, the solution for $h_1(w,z)$, the solution, correct to order e is simply

$$h_1(w,z) = \int dz' dw' G(w-w', z-z')(\chi[z'] + \Lambda[w'])h_o(w' - [z' - z_c]) \qquad (29)$$

Here, $h_o(w)$ is the value of $h(w,z_o)$ at $z=z_c$.

The next issue is what is the Greens function for the h. In one dimension and in three dimensions, the initial delta function simply propagates outward, while in two dimension, the Greens function has a wake behind the advance of the front, as one can see by throwing a pebble into a pond. In one dimension, the Greens function just splits into two propagating pulses, propagating away in each direction. That is

G = 0.5 d[(w-w') – (z-z')] + 0.5 d [(w-w') +(z-z')] (30)

In our case, the second of these delta functions will give the disturbance propagating up the characteristics to higher values of w; a solution which is clearly unphysical and must be eliminated. Keeping only the first delta function will give a propagating 'wave' only towards lower values of w, and is the proper, physically correct solution.

Doing the w' integration, we find that the zero order solution factors out. That is

$$h_1(z,w) = 0.5 h_o(w - [z - z_c]) \int dz' \left\{ \chi(z') + \Lambda(w - [z - z']) \right\} \tag{31}$$

The only issue now is what are the limits of the z' integration. To answer that, Fig. (14) provides the insight.

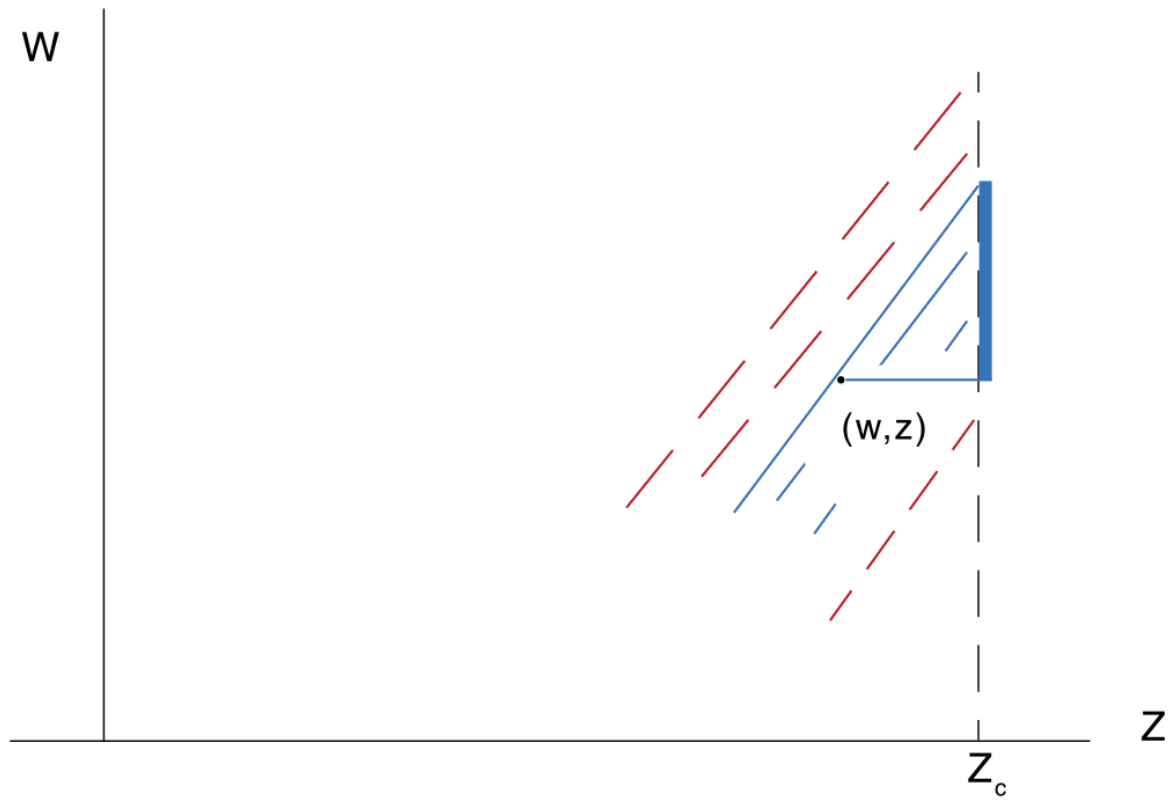

Figure (14): The wz plane. The point (w,z) is shown, as well as several characteristics. The blue shaded region of vertical axis at $z=z_c$ shows the region between z and the intercept of the characteristic passing through the point (w,z).

The issue is the propagation of energetic electrons from the point $z_c$ where they are formed, for instance the quarter critical surface, and towards the interior of the target. Hence the effect of these energetic electrons propagates down the characteristic. For the electrons at $z_c$ with w greater than the top of the blue shaded region, the characteristic misses the region where it can affect the position shown as (w,z), so it cannot have any effect there. Hence these characteristics are shown as the upper red dashed characteristics. The lower dashed red characteristics have too low an initial energy to have any effect on the electrons at (w,z). Hence, the only characteristics that can have any effect on (w,z) are those intersecting $z=z_c$ within the blue shaded region, that is the characteristics shown in dashed blue. Clearly, this means that the limits of the integral over z' in Eq (31) are from $z_c$ to z.

While the spherical part of the integral in depends on the density profile, the $\Lambda$ part can be done analytically. We find that for the planar configuration

$$h_1 = h_o \frac{(Z+1)(Z+9)}{64} \left[ \frac{1}{w} - \frac{1}{w - z + z_c} \right] \tag{32}$$

Notice that h has a singularity a w=0. However, the distribution function, which is $w^{Z+1}h(w,z)$ has no singularity. Hence using perturbation theory one can get a better approximation where perturbation theory is valid.

For the more general case, it is possible that a routine like Mathematica can solve for more general solutions., perhaps even at each step of a rad hydro simulation.

Asking Google:

Does Mathematica have efficient means of solving complex wave like partial differential equations?

One gets the AI overview

AI Overview

Yes, Mathematica possesses efficient, built-in numerical routines to solve complex, wave-like, and time-dependent Partial Differential Equations (PDEs). Using

NDSolve, Mathematica automatically applies finite element or pseudospectral methods, handling nonlinear systems, complex boundary conditions, and multi-physics modeling in high dimensions.

VI Solution using sparse eigenfunctions

Notice that for Eq (24), the rhs only operates on w, and the lhs side only operates on w, assuming Z is constant and is not a function of z. This motivates separating the equation in a product of a function of z and a function of w:

$$f_1(w,z) = \Phi(z)\Omega(w) \tag{33}$$

Then one can divide the Eq. (24) by $f_1(w,z)$ so the left side becomes only a function of z and the right side becomes only a function of w. Hence each side must be a constant with no z or w variation; a constant we call $\kappa^2$ and the two equations become:

$$\frac{d^2\Phi}{dz^2} + 2\frac{d}{dz}\left[\frac{\Phi}{k(z)\gamma(z)}\right] = \kappa^2\Phi,$$

$$\frac{d^2\Phi}{dz^2} + 2(r(z)k(z))^{-1}\frac{d\Phi}{dz} + 2\frac{d(rk)^{-1}}{dz}\Phi = \kappa^2\Phi, \tag{34}$$

and

$$\frac{d^2\Omega}{dw^2} - \frac{1}{4}\frac{1+Z}{w}\cdot\frac{d\Omega}{\mathrm{dw}} + \frac{1+Z}{4w^2}\Omega = \kappa^2\Omega. \qquad (35)$$

First let us consider Eq. (34), the spatial part, and for now neglect the spherical term. Then the solution is simple. As Φ must increase from z=0 to $z=z_c$ by several orders of magnitude at least, $\kappa^2$ must be positive. For the planar case, the solution is simply

$$\Phi \propto \sinh \kappa z. \qquad (36).$$

This equation certainly satisfies the boundary condition at z=0, and the coefficient must be such that it satisfies the boundary condition at $z=z_c$.

To continue, we examine the equation for w. It is convenient to relate this to Bessel's equation. To do so we define new dependent and independent variables, ω = kw and $\Omega = \omega^N\Psi(\omega)$. Then the energy equation can be rewritten is

$$\frac{\mathrm{d}^2\Psi}{\mathrm{d}\omega^2} + \frac{1}{\omega}\frac{\mathrm{d}\Psi}{\mathrm{d}\omega} - \left(1 + \frac{\alpha^2}{\omega^2}\right)\Psi = 0,$$

$$\text{For n} = (\mathrm{Z}+5)/8 \quad \text{and} \quad \alpha^2 = [(\mathrm{Z}-3)/8]^2. \qquad (37)$$

This is the modified Bessel's equation, and the solutions are simply the modified Bessel function $K_\alpha(\omega)$ or $I_\alpha(\omega)$. Since $I_\alpha(\omega)$ diverges for large w, the only allowed solution is $K_\alpha(w)$.

$$\Psi(\kappa w) \propto K_{\left|\frac{Z-3}{8}\right|}(\kappa w) \qquad \Omega = (\kappa w)^{(Z+5)/8}\Psi(\kappa w) \quad (38)$$

While Ψ(ω) diverges for as ω approaches zero, Ω(ω) does not. In fact, as w approaches zero, W approaches zero linearly in w. Let us say that $f_1(w)$ at $z=z_c$ just happens to be proportional to

$$f_1(w) \sim (\kappa w)^{(Z+5)/8} K_{|(Z-3)/8|}(\kappa w) \qquad (39)$$

Then for the planar case, with a single Z, we have that the solution for all z between 0 and $z_c$ is

$$f_1(w,z) \propto (\kappa w)^{\frac{Z+5}{8}} K_{\left|\frac{Z-3}{8}\right|}(\kappa w)\sinh \kappa z. \qquad (40)$$

For the spherical case, one only needs to solve Eq. (34) numerically between $0 < z < z_c$ . In either case, the coefficient of proportionality is determined so that the energy flux is as specified at $z=z_c$.

Let us now give a brief digression on integrating for the planar and spherical case. In each case, we start from z=0. For the planar case, there are two solutions near z=0, one is a constant and one has Φ increasing linearly in z. Clearly, only the one solution increasing linearly in z is physically correct. By starting the numerical integration at z=0, one can be certain that he is neglecting the unphysical solution. The spherical case is rather different. Here one can show that near the origin, the two solutions are proportional to z and to $z^{-3}$. Obviously, this latter solution is unphysical and must be eliminated. It is simplest to do this by beginning the numerical integration at z=0 with the physical solution. Both the planar and spherical cases have very different unphysical spurious solutions. Fortunately, both have the same correct physical solutions near the origin.

To continue, the problem is twofold. First the distribution function at $z=z_c$ is not given by Eq. (39), and second, the eigenfunctions do not form a complete orthogonal set, so there is no way to uniquely decompose the value at $z=z_c$ into a unique summation over the eigenfunctions. However there does seem to be a way to do what is second best and it could be the basis of forming a good approximate solution. Say the condition at $z=z_c$ is not given by a single eigenfunction, but what if it is given by two, or three, or four? Then one can get the solution for all $0<x<z_c$ by summing over the two, three or four eigenfunctions. But what if there is no small simple set of eigenfunctions that matches the $f_l(w)$ for all w? But what if instead, there is a small number that match it approximately, not for all w, but only for the w's that are most important for determining the energy flux at all w, and for other values of w go approach zero?

This is the crux of the sparse eigenfunction approach to getting an approximate solution to Eq (24). There is no guarantee that one can find such a set of sparse eigenfunctions which are a good approximation to the boundary value of the dependent variable, but for the case at hand, there does seem to be.

As we have seen, the value of the energetic energy distribution function at $z=z_c$, as a function of w, $f_l(w) \sim w^{1/4}\exp\text{-}w^{1/2}$, where the coefficient is determined so that the energy flux is a specified. In (25), this author has given some examples of determining an appropriate set of sparse eigenfunctions which closely match this boundary condition over what seem to be the main regions of w relevant for energy deposition in the target. The result for Z=3 is shown in Fig. (15). Details are given in (26), but for a related case, not in any of the author's previous work, we will go into more detail in the next section. Meanwhile, Fig. (15) shows a plot of f(w), Z=3, and an attempt to match it with 3 sparse eigenfunctions.

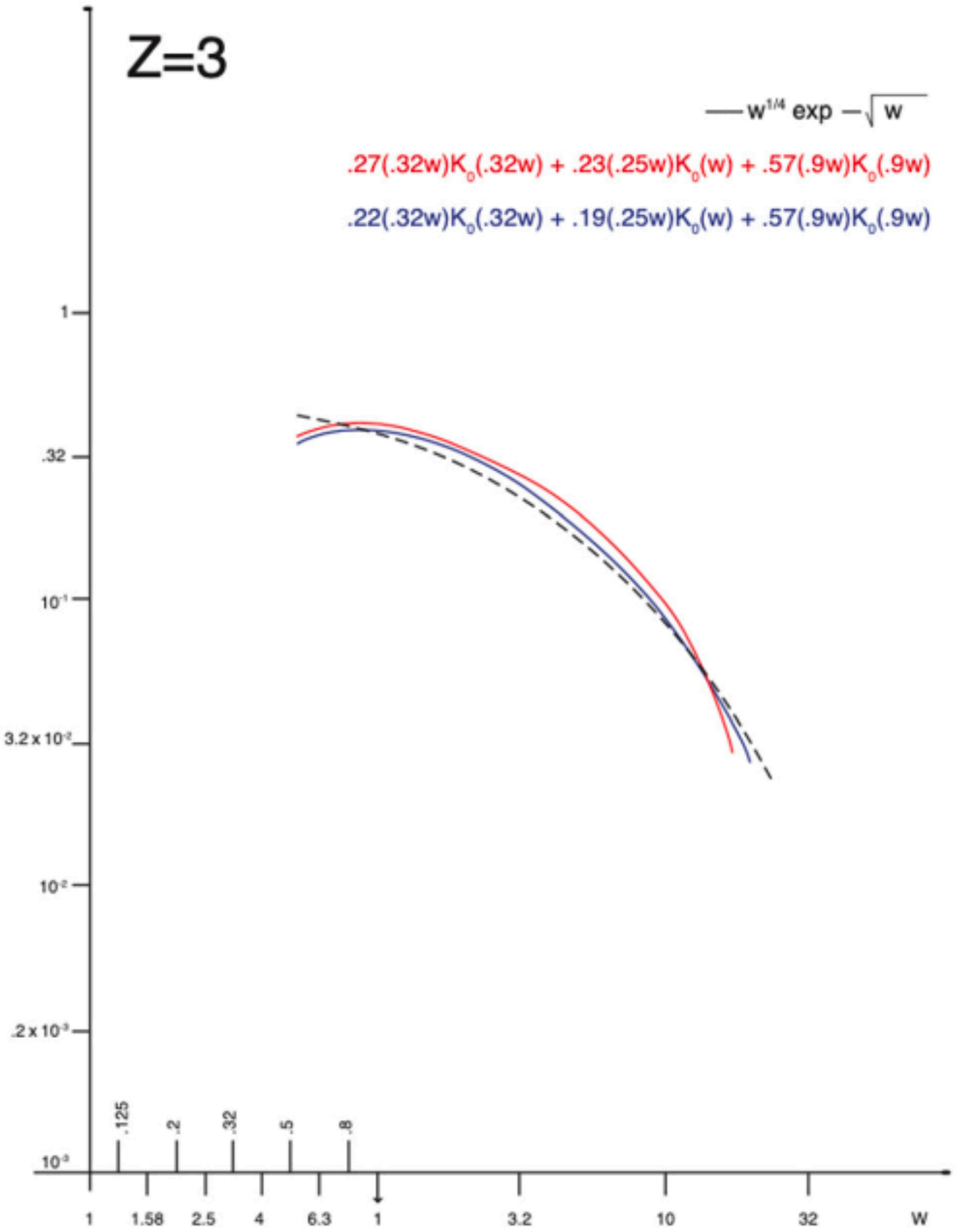

Figure (15): The distribution function $f_1(w)$ is dashed black, and 2 attempts to model it with a sum of 3 sparse eigenfunctions. For w much less than one, the sparse eigenfunctions do not match at all, but there is very little energy flux there. For much larger w's they do not match either. After all the sparse eigenfunctions decay in as exp-kw, whereas the actual distribution function decays as exp-$w^{1/2}$ . However, there are very few particles at these large values of w. This shows the nearly unavoidable approximate nature of the sparse eigenfunction approach to solving the equation.

For an unsegmented target, the solution for all values of z between 0 and $z_1$ are given by the spatial and velocity dependences of the summation over the 3 sparse eigenfunctions like those given in Eq.(40). The coefficient of each sparse eigenfunction is given so as to match the boundary condition at $z = z_c$.

The only thing remaining is the case of a segmented target with a variety of Z's. The author examined one case of this motivated by a n experiment by the URLLE/LLNL group on the NIF laser (27) and the result seemed to give reasonably close agreement with the experiment (26). Recall that the equation for $\Phi$ has no Z dependence, so the several values of $\kappa$ (3 of them in the above case) for all $0 < z < z_c$ are determined entirely by the distribution function at $z=z_c$..

However, the functions of w take a jump at the interface between, say $z_1$, where Z jumps from $Z_1$ to $Z_2$. We treat this by asserting that for each sparse eigenvalue, the there is no discontinuity in its energy flux as it passes from one region to another.

Hence starting from z=0, one integrates the equation for Φ, with the boundary condition at z = 0 given by Φ = 0 and d Φ /dz = A, up to the next segment. Here A is a constant to be determined. At the next segment, the coefficient of the sparse eigenfunction most likely changes so as to conserve energy flux. Then one keeps going through as many segments as there are until one reaches $z=z_c$. At this point one knows what the coefficient of the sparse eigenfunction is. Since the equation for $f_l$ is *linear*, the dependence of the calculated sparse eigenfunction depends *linearly* on A. At this point, it is a simple matter to solve for A so that the sparse eigenfunction at $z=z_c$ satisfies the boundary condition.

Finally, one must admit that as specified so far, the decomposition of the value of $f_l(z_c)$ is not unique. Different sparse eigenfunctions used in the decomposition, will give slightly different answers, and none will be perfect matches. However, if the sparse eigenfunction decomposition is reasonably close, to the actual $f_l$ in the region of w which is of interest, the approximation should be reasonably good. It is likely that the choice of sparse eigenfunctions could be improved, by for instance choosing their coefficients with a least squares fit for instance.

## VII. Other High Z sparse eigenfunctions

The basic equations for $f_l$, Eq. (12) suggests another approach to the sparse eigenfunction method for solution. For high Z one can consider neglecting the second derivative term on the right-hand side. This may be relevant for the high Z plasma generated on the inner wall of the hohlraum in experiments like those done on NIF.

Of course, by neglecting this term, it is no longer possible to satisfy both boundary equations in w, the condition at w=0 *and* the condition at w equals infinity. We will discuss this more shortly.

As we have seen, the actual solution of Eq. (12) in terms of modified Bessel functions of the second kind, approaches zero linearly as w approaches zero, and approaches zero exponentially as w approaches infinity. Neglecting the second derivative term, the 2 equations (34 and 35) would be replaced with

$$\left\{\frac{4}{1+Z}\right\}\left[\frac{d^2\Phi}{dx^2}+2\frac{\mathrm{d}}{\mathrm{d}z}\left(\frac{\Phi}{k(z)r(z)}\right)\right]=\vartheta^2\Phi \tag{41}$$

$$-\frac{\mathrm{d}\Omega}{\mathrm{d}w}+\frac{\Omega}{w}=\vartheta^2 w\Omega \tag{42}$$

And of course, these equations and the others in Section VI are related by

$$\left\{\frac{4}{1+Z}\right\}\kappa^2=\vartheta^2 \tag{43}$$

assuming Z is constant.

Notice that now Z can be a function of z, since it appears only in the spatial equation and no longer in the equation for w, i.e. energy.

It is not difficult to see that the solution to Ω is

$$\Omega = A\,(\vartheta w)\,exp-(\vartheta w)^2/2 \equiv A\xi exp-\xi^2 \tag{44}$$

Here A is an arbitrary constant used to normalize the expression in some way to be specified shortly.

Of course, we recall that the solution of the actual equation Eq. (35) for constant Z is proportional to

$$\Omega=(\kappa w)^{(Z+5)/8}\Psi(\kappa w) \quad \text{where} \quad \Psi(\kappa w)\propto K_{\left|\frac{Z-3}{8}\right|}(\kappa w) \tag{45}$$

Notice that Eq. (45) for Ω approaches zero linearly in w, as does Eq.(44). However, as w approaches infinity, Eq (44) approaches zero as a Gaussian, while Eq(45) does so as an exponential. Hence while the two solutions may approximate each other for small w, they cannot for very large w.

To see how valid the approximations we made are, it is simple to compare Eq. (44) with Eq (45) for a variety of Z's. To do so we use Eq. (44), but for the ξ variable, so that in terms of this variable, there is no dependence on Z. Then expressing Eq. (45) for the Besssel function solution, also in terms of ξ, we get a different result for each Z. Figure (16) shows the result. Each expression is normalized so that each curve maximizes at x = 43.

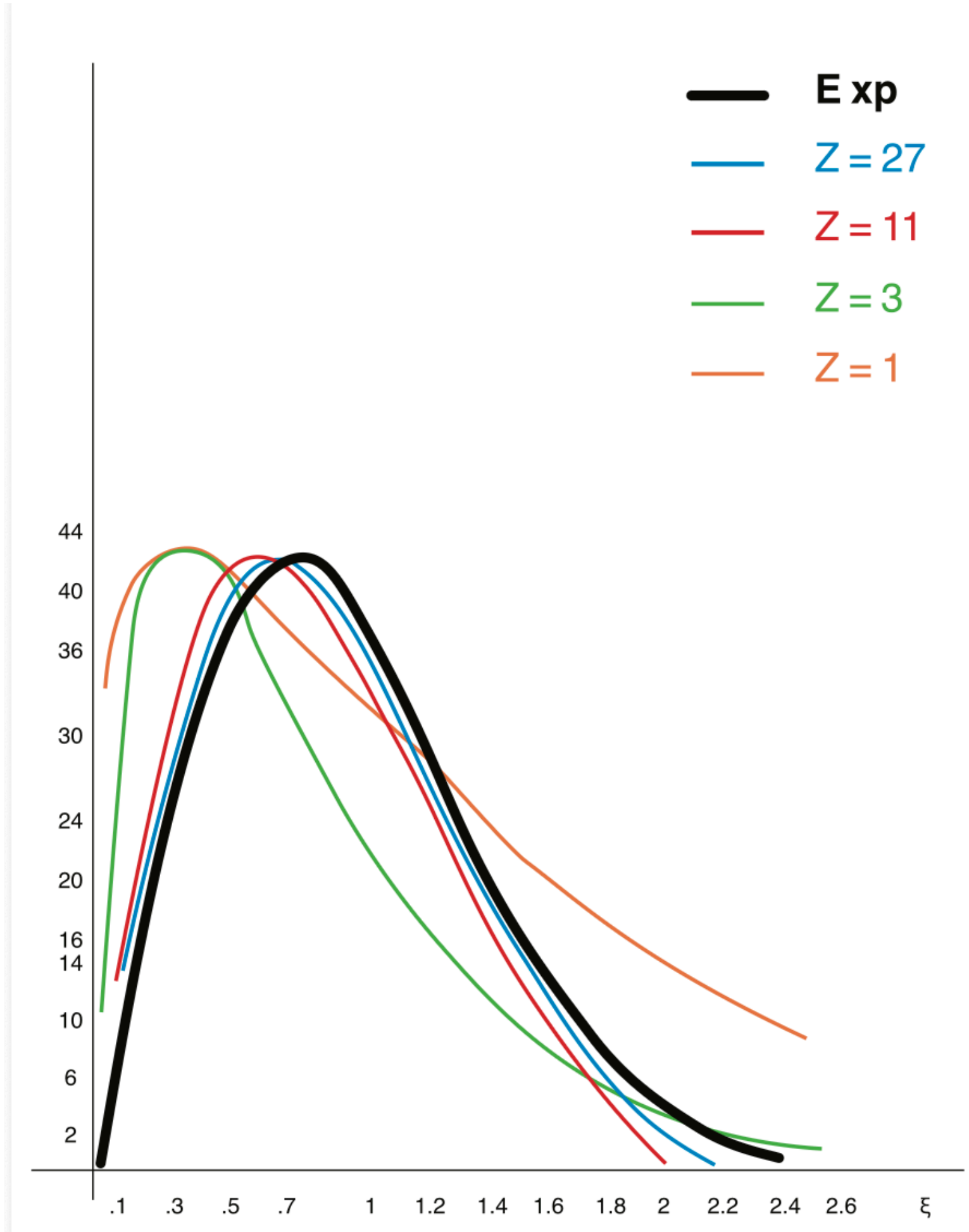

Figure (16): A plot of the exponential solution, Eq. (44), heavy black line; the Bessel function solutions for Z=27, blue line; Z=11, red line; Z=3, green line, and Z=1, orange line. Clearly the exponential approximation is very good for Z=11, and 27. Perhaps it is barely passable for Z=3, and is quite far off for Z=1.

Continuing, we next examine whether the $f_1(w,z_c)$ can be reasonably approximated with a small set of the new sparse eigenfunctions. Recall that

$$f_1\,(w, z_c) = Aw^{.25}\,exp - w^{.5} \qquad (46)$$

Here A is determined by the energetic electron energy flux. Taking A =1, Figure (17) is a plot of $f_1$, the heavy black line. Also shown, as the heavy red line is a plot of the sum of 3 sparse eigenfunctions, one maximizing at w=2, another a w=8 and a third at w=20. Between w =2 and 30, the match is reasonably close. Between 0<w<2, and w>30 it is poor. However, the particles with w<2 have very little of the energy flux; and for w>30, there are very few particles.

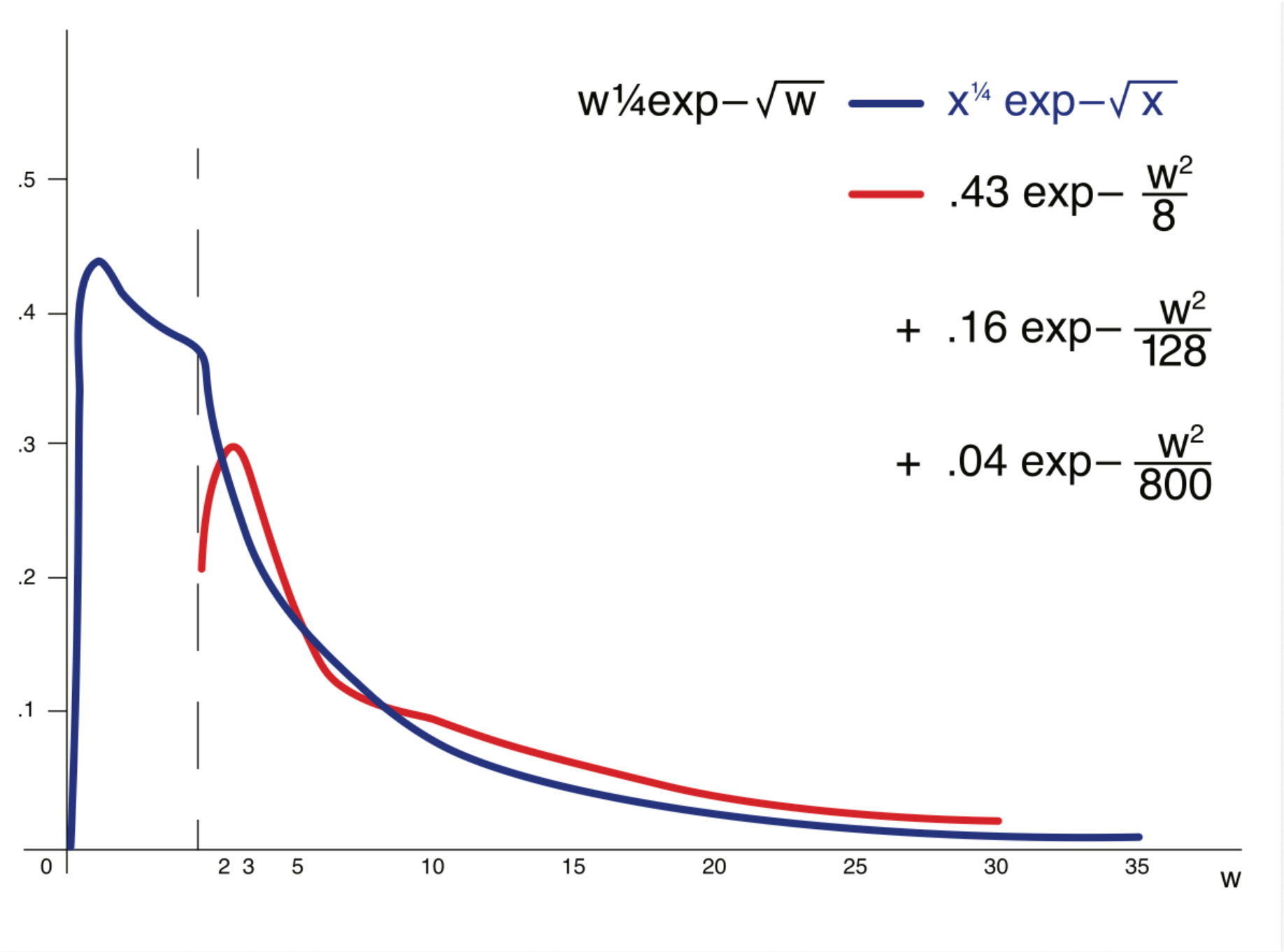

Figure (17): An approximate match of $f_1(w,z_c)$ with 3 sparse eigenfunctions.

Hence it seems that for large Z's the one can find a reasonable approximation for the chosen value of $f_1(w,z_c)$.

Notice that these new sparse eigenfunctions allow Z to be a function of radius (i.e. z). For instance, for the planar case, one can easily integrate the spatial equation, Eq. (41) from z=0 to $z=z_c$. If a WKB approximation is valid, one simply has a hyperbolic sign for each sparse eigenfunction

:
$$\Phi(z) = \int_0^z dz' \, sinh\vartheta \sqrt{\frac{1+Z(z')}{4}} \qquad (47)$$

Finally, one must normalize the above function so that $F(z_c)$ has the proper value at $z=z_c$ for each sparse eigenfunctions.

It is interesting that Ref (22) also considered high Z effects. In fact, in their initial simple example, they chose a Z = 100. Well, who knows, maybe somebody will someday produce a fully ionized Fermium plasma. However, they also considered what they thought of as a plasma produced on the wall of a hohlraum filled with a helium gas. They envisioned this plasma produced as having a spatial variation of Z going from 2 to 36. Possibly theory developed in this section could be helpful to them if there are energetic electrons involved.

## VIII. Fuel preheat from energetic electrons in the tail of the Maxwellian.

Laser plasma instabilities (LPI's) are hardly the only source of energetic electrons which may preheat the fuel part of the fusion target. There is also the tail of the Maxwellian to consider. Depending on plasma temperature and density these energetic electrons may only be weakly coupled to the local background plasma and may travel a considerable distance before they are thermalized. This has hardly escaped the attention of the laser fusion community. In fact, as far as this author can discern, there are many more studies and attempts to describe this phenomenon than there are attempts to study the effects of LPI's on fusion gain.

Most of these studies attempt to develop an approach to the entire electron distribution function which describes both the thermal and energetic electrons. As with the case of the laser plasma instability, this work takes a different approach. The main assumption is that there are only a small number of electrons that are decoupled. They do not interact with each other but only interact with the background electrons and ions. Furthermore, as in the case of the LPI's, we assume that they thermalize in a time scale virtually instantaneous as compared to the fluid motion. Hence, we solve the Fokker Planck equation for these energetic electrons only in space and assume the background has no time dependence.

This brings up two immediate dilemmas which do not occur in the LPI case. First, where does one draw the line between the thermal and energetic electrons, and second, since it is a spatial problem, where is the spatial boundary? For the instability cases considered, the spatial boundary is obviously at the position of the quarter critical density, and the energy boundary is simply considering separately the thermal (~2.2 keV) plasma, and the instability generated plasma (~50keV).

However, we begin a description of the transport properties of a plasma and how they can be calculated. The gold standard in describing the collisional and transport properties of a plasma is the work of Braghinskii (45). This work gives the functional dependence of every transport coefficient in terms of the plasma properties. Also, it gives what are generally considered to be the most accurate numerical coefficients which multiply these functional dependences to get accurate local transport coefficients.

However, there is a shortcut which has been frequently used, certainly by this author and many others. Using a Krook model, one gets the parametric dependences of the transport coefficients correct, but not the coefficients. Hence it is very tempting to use the Krook model to derive the nonlocal transport but then correct the coefficients so that they agree with the Braghinskii calculation.

The reason to do this, is that it is relatively simple derive a nonlocal theory for the entire plasma using a Krook model, but to this author's knowledge, there is no relatively simple way to do an analogous nonlocal extension of the Braghinskii calculation. Once one has the nonlocal theory from the Krook analysis, the spatial boundary conditions, and the dividing energy between those electrons that deposit their energy locally, and those that deposit nonlocally; it is a relatively simple matter to switch the nonlocal part to a Fokker Planck analysis.

Hence let us consider a Krook model for the distribution function. Take the simplest steady state expression of the Krook model for the $f_o$ and $f_1$ (1).

$$\frac{V}{3}\frac{df_1}{dx} = -\nu_e(f_o - f_M) \qquad V\frac{df_o}{dx} = -(\nu_i)f_1. \tag{48}$$

In Eq. (48) we neglect the effect of the electric field on the motion of the energetic electrons, for the same reason as we did in the LPI case. Namely the electric potential across the plasma is roughly the maximum temperate, while the energy of the relevant electrons is considerably greater. Of course, this assumption is more difficult justify than for the case of LPI generated electrons. However, we will see shortly that this may not be as serious an issue as one might expect.

As in Section III, these can be combined to form a single second order equation in x equation for $f_1$. Since there are no velocity derivatives in Eqs (48), one can multiply this second order equation by $(1/2)mv^3$ and one has a second order equation for $q(x,\Xi)$, the electron energy flux at each energy $\Xi$.

$$-\frac{1}{k}\frac{d}{dx}\frac{1}{k}\frac{dq(x,\Xi)}{dx} + q(x,\Xi) = q_L(x,\Xi) \tag{49}$$

$$q_L(x,\Xi) = -\frac{K_{sp}}{\Lambda\beta(\infty,Z)} \frac{\Xi^4\left[\left(\frac{\Xi}{T}-\frac{Z_4}{Z_2}\right)\frac{1}{T}\frac{dT}{dx}+\delta\varepsilon\right]\exp-\left(\frac{\Xi}{T}\right)}{T^{3/2}P\left(\frac{\Xi}{T},Z\right)}$$

(3 (50)

The quantity $q_L$ is the local expression for the energy flux as generated by the various gradients using a Krook model, but normalized so that the integrated result gives the classic value for electron thermal transport. The quantity k is closely related to the reciprocal of the energy dependent electron mean free path, and $K_{sp}$ is the electron thermal conductivity as derived by Braghinskii:

$$k = \frac{1}{V}\sqrt{3(\nu_e+\nu_i)\nu_e} \equiv \frac{1.15\times10^{-13} n\Lambda\sqrt{1+Z}}{\Xi^2} \tag{51}$$

$$K_{sp} = \frac{9.7\times10^8\gamma}{Z}\frac{\text{ergs}}{(\text{cm})(\text{sec})(\text{eV})^{7/2}} \quad \gamma = 13.3\frac{Z+0.24}{Z+4.24}. \tag{52}$$

The quantity $\delta\varepsilon$ is the perturbation to the electric field generated because some of the energetic electrons do not deposit their energy locally; it is calculated in Refs. (14,24). The functions β and P are various normalizing factors, details are in (14,24).

Clearly for low energy electrons, the mean free path is short, and k is large. Hence for these electrons $q \approx q_L$. For higher energy electrons, the second derivative term, i.e. nonlocal effects, becomes important. If one does a purely Krook calculation, one simply solves Eq. (49) in x for all Ξ, integrates over Ξ, and uses this expression for thermal electron energy flux this in the rad hydro fluid equation. A great deal of our earlier work did just this (9-14).

However, as we have discussed the Krook approximation greatly overestimates the preheat in a laser fusion direct drive configuration. Our aim is to use a classical approach for the low energy particle, and a Fokker Planck approach for the few high energy particles on the tail of the Maxwellian.

The next issue how does one separate the two parts of the distribution function. In our earlier work (9-12), we attempted to define this energy in terms of local gradients. However, on running the rad hydro code, the gradients were bouncing around a great deal and that approach proved not to be viable. We found that a much better approach proved to be using integrals instead of differentiating.

Our definition is now that that the critical energy differentiating the low and high energy parts of the distribution function, $\Xi_m$ is calculated by the relation in Eq. (53)

$$\int_{r(T_e/2)}^{r(T_e)} drk(r, \Xi_m) = 1 \qquad (53)$$

Here $T_e$ is the maximum temperature in the spatial temperature profile at the time of the calculation, and $r(T_e)$ is the position of this maximum. Then $r(T_e)/2)$ is the position, going to higher density, where the temperature is half the maximum. That is that in one mean free path, the critical electron moves from the maximum to half the maximum temperature. Higher energy electrons, of course go even further.

The next issue is to determine the spatial boundary, that is the spatial position where the energetic electrons begin and what their distribution function is there.

We choose the position of origin of the energetic electrons as the position where the Krook calculation of the energy flux maximizes, defined as $r_{max}$; and we choose for the distribution of energetic electrons at that position, the local Krook calculated distribution function for $f_1$.

This initializes the distribution function $f_1$ in just the same was as it was initialized for the laser plasma instability production of energetic electrons. The Fokker Planck calculation is then done in the same way. There are some differences regarding the definition of w and z. Here

$$dz = -\sqrt{3}\frac{2.7 \times 10^{-13} n(x)\Lambda}{T_e^2} dx \quad \text{and} \quad w = \left[\frac{\Xi}{T_e}\right]^2 \qquad (54)$$

Let us look at a typical profiles for which we applied this theory.

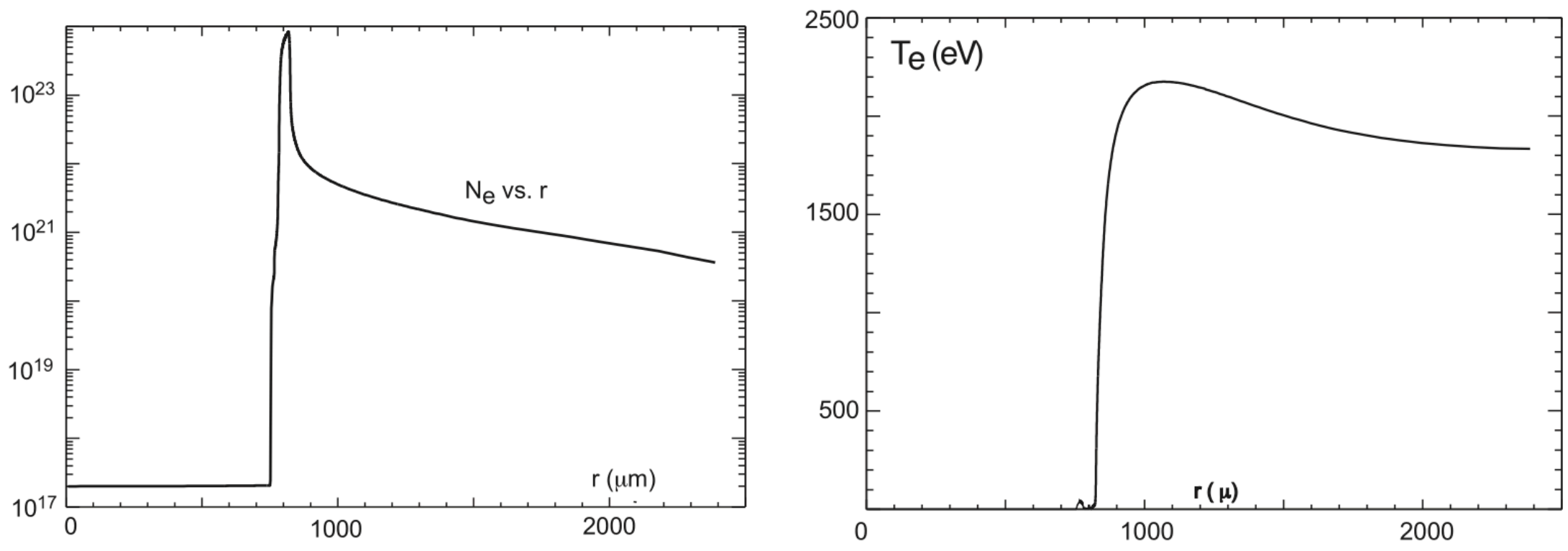

Figure (18) The spatial temperature and electron density in $cm^{-3}$ profiles for a typical optimized NRL calculation of a laser fusion target implosion at a time of 15 ns into the implosion. The maximum temperature $T_e$ is about 2.3 keV,

For the profiles above, $\Xi_m$ was about 10keV and the maximum $\Xi_m$, the energy at which there is no significant energy flux is about 20keV. Correspondingly, the range of w is from about 100 to about 400. The maximum temperature is located at about 1000μm, the position where the temperature drops to about half its value is at about 900μm, and $r_{max}$ is at about 950μm.

While the approximations we have made may be only marginally satisfied near the position of maximum Krook energy flux (i.e. z of approximately 900μm), moving inward slightly, the conditions are well satisfied, since the temperature profile has such a sharp drop in temperature as one moves inward from the maximum.

Solving the Fokker Planck Equation for the for the distribution function for $0<r<r_{max}$ and $\Xi_m < \Xi$, solves for $f_1$ in that region. Integrating over $\Xi$ from $\Xi_m$ to infinity, one determines the nonlocal component of the electron thermal energy flux as a function of x. In other words,

$$q(x) = q_L(x, \Xi < \Xi_m) + \int_{\Xi_m}^{\infty} d\Xi q(x, \Xi). \tag{55}$$

Here $q_L(x,\Xi<\Xi_m)$ is the integral of the local energy flux from an energy of zero to a maximum of $\Xi_m$. In terms of the normalizing factors of Eq (50 and 52) it is

$$q_L(x, \Xi < \Xi_m) = -\frac{K_{sp}T^{5/2}}{\Lambda}\frac{dT}{dx}\frac{\beta(\Xi_m/T, Z)}{\beta(\infty, Z)}. \tag{56}$$

Not only does Eq. (55) show that there is an additional energy flux caused by the nonlocal deposition of some of the more energetic electrons, there is also a reduction of flux given by the fact that the integral over the local $q_L$ no longer goes from zero to infinity, but goes from zero to $\Xi_m$. This is the way this current theory describes flux limitation.

Let us continue by imaging what the distribution function looks like at for instance r=500μm. First there is the central low temperature region, with energy of, let's say a few electron volts Then there is a much higher energy component, with energy between ~ 16 keV (i.e. $\Xi_m$) and 40 keV. An image of such a distribution function is shown in Fig (19):

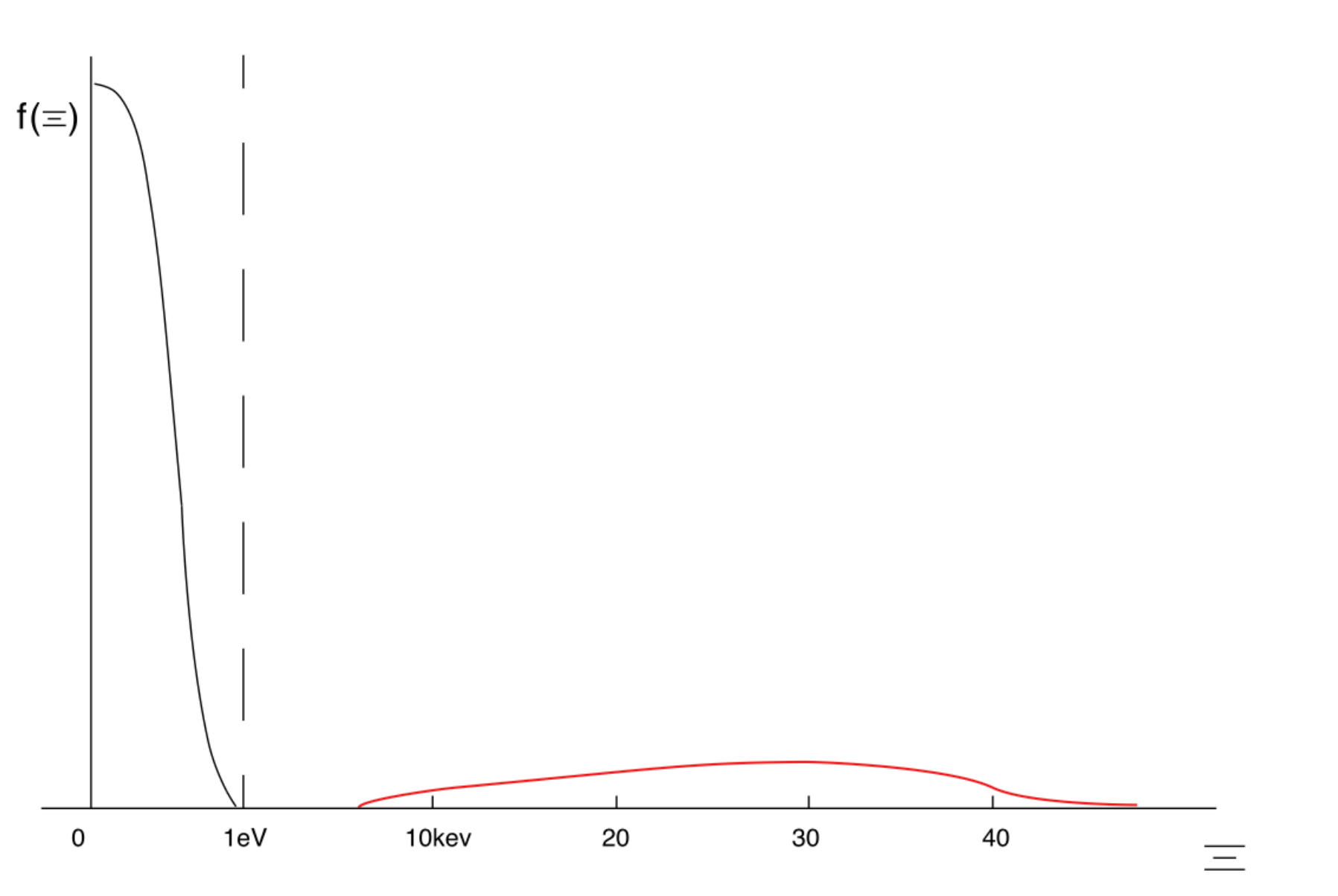

Figure (19) A rough cartoon of what the electron distribution function looks like at ~ 500μm for the profiles of Fig (18). Notice the change in scale on the horizontal axis at ~ 1eV. Surely no localized perturbation theory at around 500μm could have predicted this.

Notice the general shape of the distribution function. If ought to be obvious that doing some sort of localized perturbation theory, as several researchers have attempted, it would never result in such a distribution. In fact, the high energy halo in Fig (19) does not arise from any localize process, but it originated at ~ 950μm propagated to 500μm.

To solve the Fokker Planck equation with a sparse eigenfunction technique, we must first see if we can get a set of sparse eigenfunctions that approximate the Krook calculation of the distribution function, $f_l$, at $r_{max}$. This was discussed in Refs (24,25). Figure (20) is plot of $f_l$ at $r_{max}$ as a function of w, and on the same plot, is a function of an approximation to it by a single sparse eigenfunction.

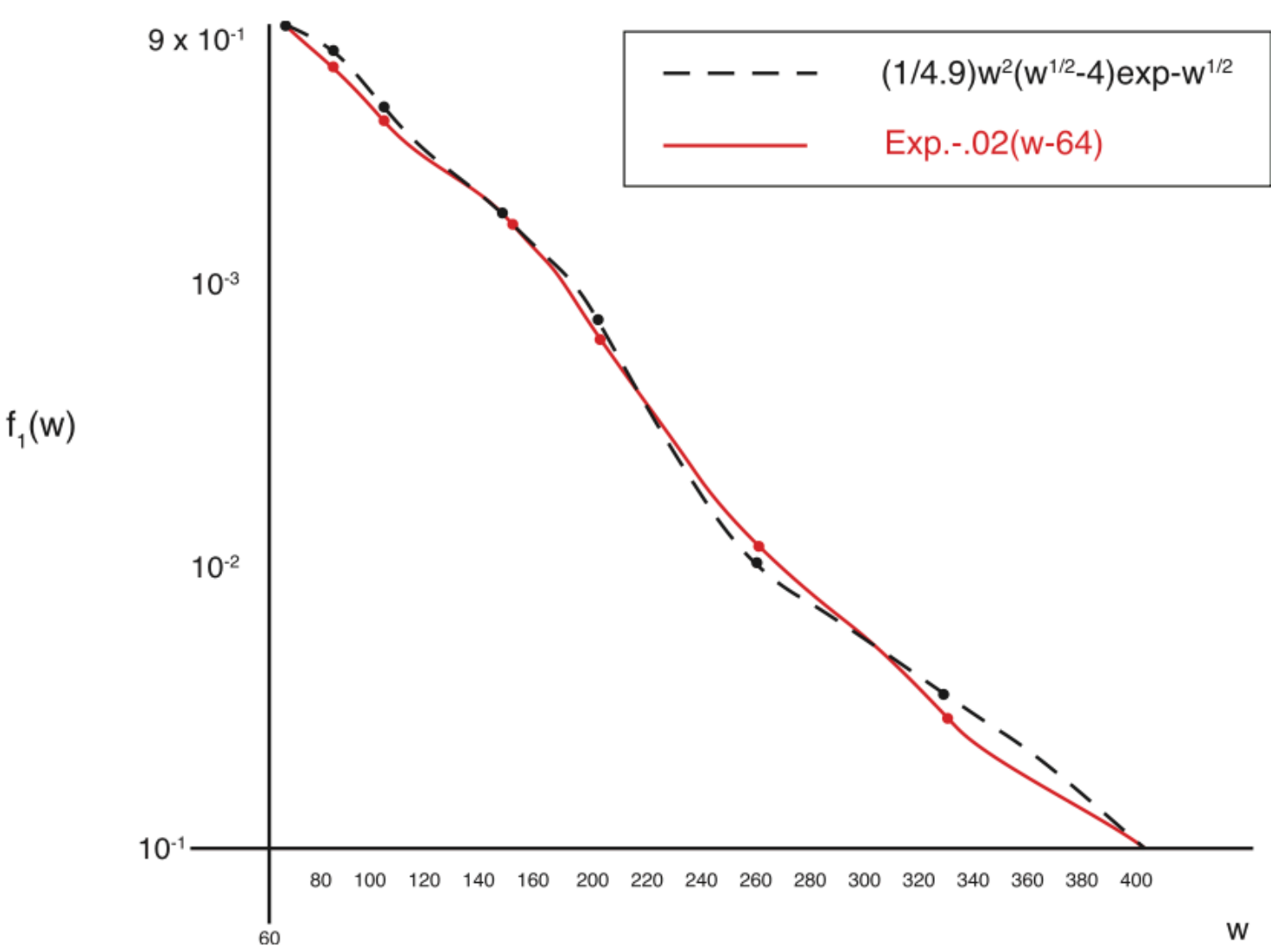

Figure (20): A plot of $f_1$ at $r_{max}$ for w between 60 and 400 for the density and temperature profile shown in Fig (18). (dashed black curve), as well as an approximation to it with a single sparse eigenfunction (red curve)

With this information, we can solve for the total electron thermal energy flux in three ways, first a pure Krook calculation, second a Fokker Planck calculation using the characteristic method, and third, the same calculation using sparse eigenfunctions. One example of the results of such a calculation, for the density and temperature profile, but for Z=1, from Ref (25), are shown in Fig. (21).

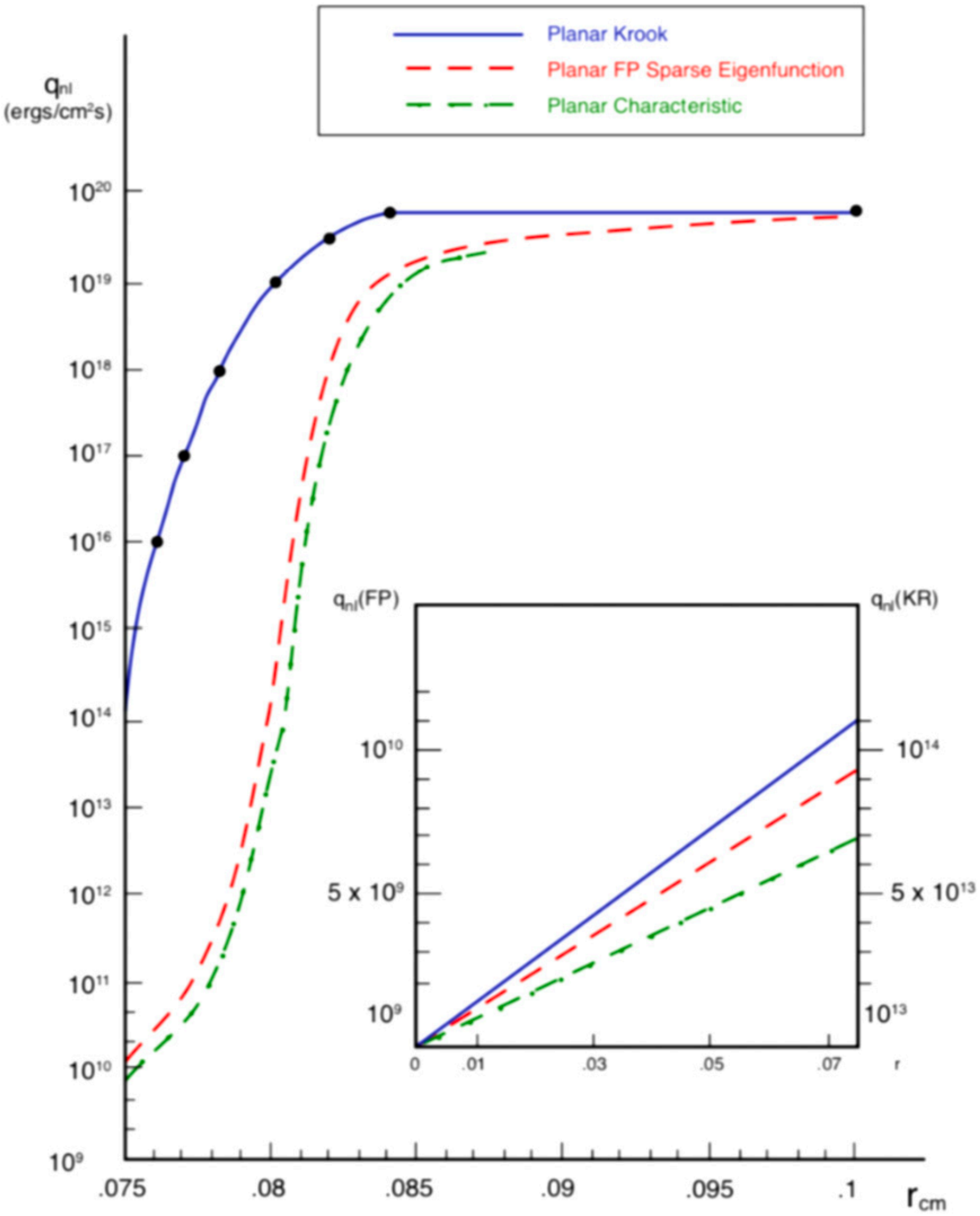

Figure (21): A calculation of the results for the thermal energy flux for the hot electrons on the tail of the Maxwellian distribution, mostly from the hottest part of the plasma around r ~1000μm. The figure shows 3 different methods of calculation, pure Krook, FP by sparse eigenfunctions, and FP by characteristics.

Notice that the pure Krook calculation, basically the Schurtz model (1), predicts four orders of magnitude more preheat than do either of the Fokker Planck calculations. No wonder the approaches by many different researchers using basically the Schurtz model have included a variety of methods to reduce the strong preheat that it predicts. Not only that, but our two methods of doing a Fokker Planck calculation, characteristics and sparse eigenfunction, give reasonably similar results. Furthermore, they give results much more optimistic for the future of laser fusion.

While it is difficult to claim that our model can accurately predict the preheat, the author does claim, with humility, that it is a very significant advance in the theory of target preheat by decoupled energetic electrons, either from an instability or from the tail of the Maxwellian. The author hopes he is being neither overly modest, nor overly immodest.

## IX. Some Digression

The digressions included here include pointing out some other aspect of the problem which we have considered, but have not discussed here, some other things to suggest which would be rather easy to do, and a few more suggestions that would not be so easy to do but may be important. They are separated into subsections denoted with capital letters.

### A: The Coulomb logarithm, the Legendre expansion

In Ref (24) was a discussion of the coulomb Logarithm, $\Lambda$. This is usually defined as the logarithm of the ratio between the distance of closest approach of the colliding particle, and the maximum distance, where the particles can interact. The velocity used in determining the distance of closest approach is usually taken as the electron thermal velocity., and the distance of closest approach is usually taken as the deBroglie wavelength for this particle, and the maximum distance usually taken as the Debye length, the distance over which individual charges are shielded. But this is only reasonable for a Maxwellian plasma. The plasma over a large part of its orbit, Fig (19), is like a two component plasma; not only that, but it is the development of the high energy component that we are interested in. Hence a more reasonable approximation is to use the closest approach for the typical particle in the energetic part, not the thermal part of the distribution. Its deBroglie wave length is much shorter than that for the low temperature plasma. Hence $\Lambda$ for these energetic particles should be much larger than the local $\Lambda$ for the cooler plasma. This is discussed further in (24). One example is the calculation of the $\Lambda$ for energetic electrons of the density profile in Fig (18) (i.e. energy of order the maximum temperature). This is shown in Fig (23).

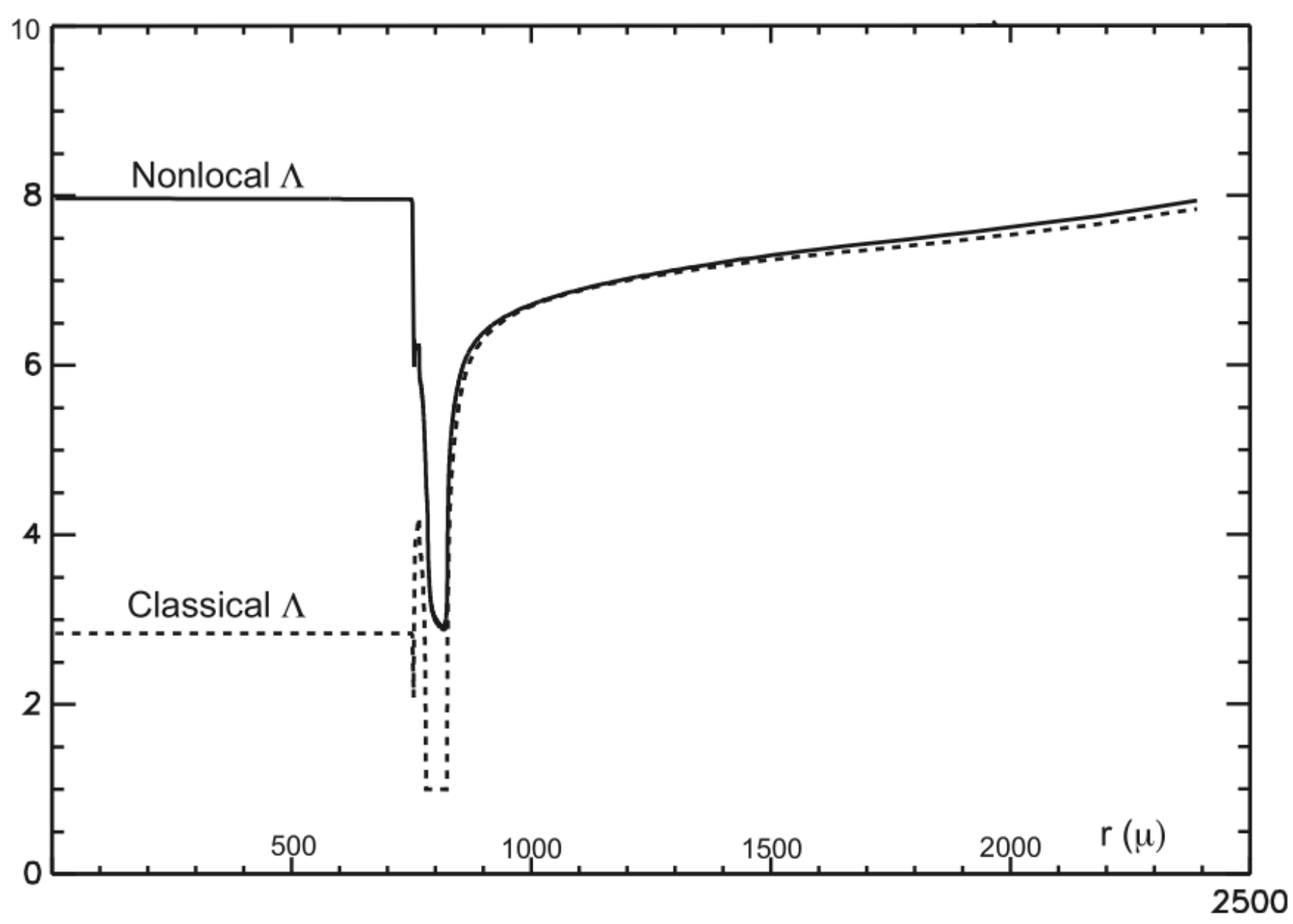

Figure (22) A comparison of the nonlocally calculated Λ and the classical result, as calculated in (24) for the density and temperature profile shown in Fig. (18)

Reference (26) also discussed the effect of truncating the Legendre approximation with only 2 terms. What it found (by making several approximations) was that if one used the sparse eigenfunctions to calculate the solution of the Fokker Planck equation, the addition of the $f_2$ term had the effect of reducing the κ's. In other words, the energetic electrons penetrate further into the cold plasma than the 2 term Legendre expansion predicted. However, this reduction in k was strongly dependent on Z. For all Z, the 2 term estimates are qualitatively correct but are much more accurate at higher Z. However, for Z = 1, there were order unity corrections; but for Z as small as Z=3, the corrections were much smaller. Hence for Z equal to or greater than three, the two-term expansion should be very accurate. However, for Z=1, it may become necessary to use higher order expansions if more than qualitative accuracy is required.

B. Try to incorporate this into a rad hydro code

One thing about the characteristic solution to the Fokker Planck equation, for either an instability or the decoupled tail of the Maxwellian, is that it gives a very simple formula for the preheat, Eqs (20, 21). This is just an added heating term to insert in the energy equation. The author is almost certain that it would be rather simple to incorporate in a rad hydro code. While the sparse eigenfunction approach would be more complicated to program than just adding a relatively simple formula, it would not be that complicated either. With both coded up, there would be two independent calculation, and hopefully they would not give very different results, as was the case in Fig. (21) .

At NRL, we had planned to do this, but between the pandemic, Denis Colombant's health issues, Andy Schmitt's retirement, and the termination of the NRL program, we never were able to do so. It seems to me that it would be quite easy to do for LLNL, URLLE, or even smaller some university programs that have access to rad hydro codes which they can tinker with. One important question which the author feels it could give a preliminary answer to is:

Does the laser fusion configuration really have to operate below LPI instability thresholds?

We have seen that the processes studied gives much less preheat than the earlier Krook models. Maybe the laser fusion configuration can give increased gain if it is above, but not too far above the instability thresholds. Probably the best instability to initially investigate this the absolute Raman side scatter, because at least from what we know so far, there is little doubt that the temperature of the energetic electrons will be ~ 50 keV. From the LLNL/URLLE results, i.e. Fig (6), we have data on what the energetic energy flux for the energetic electrons will be at two point above threshold. We know that at threshold there should be no instability, so we have already 3 points on the curve of energy flux versus ratio of irradiance above threshold value (i.e. a curve also like Fig (5)).

Using this one can imagine first doing a series of runs of implosions without modeling the energetic electron energy deposition. Presumably, these initial calculations will show higher gain for higher irradiance, and one can plot the curve of gain vs irradiance. Then redo the calculations, including the formulas of the energetic electron energy flux listed in Section (II). Presumably the gain will be reduced. However it may allow the region of reasonable gain to extend to higher irradiances, and these may have higher gain. This is demonstrated by the simple graph in Fig. (23).

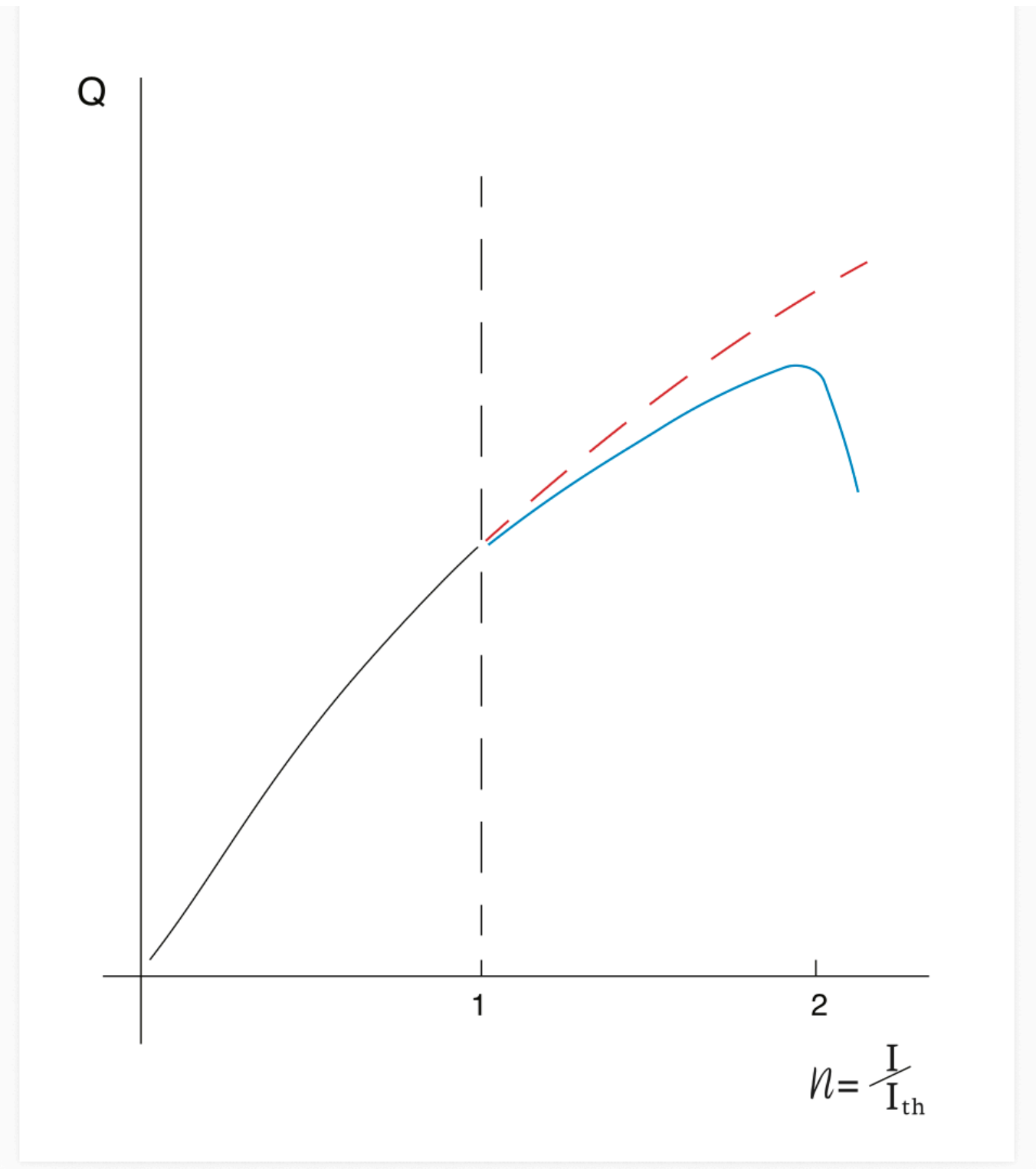

Figure (23): Two possible curves, from rad hydro simulations of the fusion gain (schematically) vs the ratio of laser irradiance to threshold irradiance (for instance) for the absolute Raman side scatter instability. At the threshold irradiance there is a break. The dashed red curve shows the gain without considering the instability. The blue curve shows the possible effect of the instability, calculated as described here. Notice that the blue curve is below the red curve, but nevertheless, the gain maximizes in the region of instability.

C. Spherical and Braghinskii

Virtually all the work we have done so far on this project in the NRL program has been for planar plasmas. For the case of solution by sparse eigenfunctions, this means that that the spatial solution for the sparse eigenfunctions are simply hyperbolic signs. Since for the sparse eigenfunctions, the spatial and velocity space solutions are handled separately, in either case, the sparse eigenfunction are the determined by the distribution function at $z_c$, and are the same whether the configuration is planar or spherical. Thus the κ's are determined only at $z_c$ and are the same whether the plasma is spherical or planar. Hence if the plasma is spherical, one simply solves Eq (34) for $\Phi(z)$, but with the spherical terms included. In the NRL program, we only did this for a single case. It was for a plasma where the energetic electrons were formed by a $2\omega_p$ instability. The temperature of the hot electrons was taken as 35 keV. Figure (24) shows the comparison of the spherical and planar results.

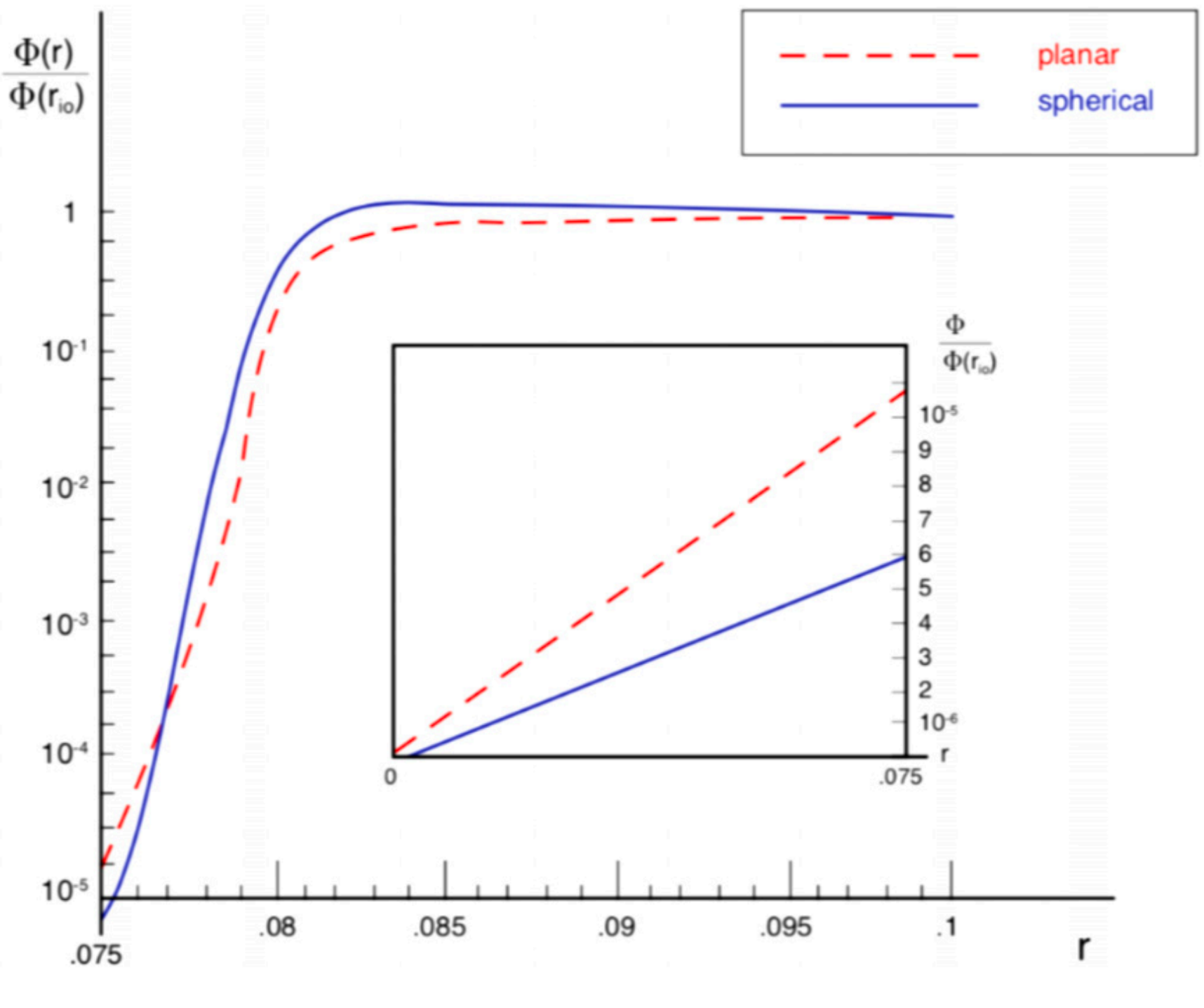

Figure (24) A comparison of the relative energetic electron heat fluxes for the a density profile like that given in Fig (18), for the spherical and planar case. The assumed instability is the $2\omega_p$ instability and the temperature (35keV) and fluxes are given in Fig (5).

The results were not very different, at least for this single case considered, but the fuel heating was smaller for the spherical case. We called this a centrifugal barrier, analogous to the centrifugal barriers in atomic physics. This is certainly a positive result for laser fusion, assuming it persists as more cases are studied.

Regarding the locally decoupled tail of the Maxwellian, we first took a Krook model for the collisions and calculated the functional dependence of the local transport. The next step was to replace coefficients with more accurate ones as calculated by Braghinskii (45) (the functional forms are the same in Krook and Braghinskii, but the coefficients are different). Then we separated the distribution function into local and nonlocal components both the configuration and energy space. The local components gave a reduced thermal flux; it is the way the theory models flux limitation. Then it used a Fokker Planck theory to calculate the nonlocal behavior of the more energetic electrons but neglected both the electric field and energy diffusion for these electrons.

It would have been better had we neglected the first step, and calculated the local distribution function initially as Braghinskii did. Also, it would have been better if had we calculated the effect of the electric field and electron energy diffusion in the nonlocal part, especially near the energy and spatial separation points. These effects would only be important in the high temperature region near the spatial point, and the energy point, where the nonlocal electrons formed. Clearly these are problems for future scientists who decide to study this method for treating nonlocal transport for energetic electrons in the tail of the Maxwellian.

D A possible FP code solution of the plasma just beyond the quarter critical density

In Sec (II) we showed that by both experiment and PIC simulations, the absolute Raman side scatter instability generated electrons leave the quarter critical surface and move inward, with their motion confined to about a $45^0$ angle with the normal to the density gradient. However, this was not the distribution we assumed as a spatial boundary condition for our calculation of the Fokker Planck solution for the deposition of the energetic electrons. Instead, we assumed a drifted Maxwellian, or more accurately assumed that the electron distribution was averaged over a constant energy surface as much as it could. However, the distribution function could not become isotropic. The energy flux had to be conserved and as the collisions of the energetic electrons with the background plasma, both with background ions and electrons, were mostly energy conserving. Hence just within quarter critical surface, the energetic electron density will build up, while conserving, or nearly conserving the total energetic energy flux to just what was produced by the instability.

This seems like a problem that could be further investigated by particle simulations, either PIC or a numerical solution of the Fokker Planck equation. Here we concentrate on the PIC simulation. The configuration would be one dimensional in space, a planar configuration would likely be a reasonable approximation, but three dimensions in velocity space. Within this region, there would be a background plasma with low temperature electrons and immobile ions with a particular Z. At the left-hand boundary, which might be, perhaps, at say $0.25n_c$, there would be an inward flux of energetic electrons, Maxwellian in energy at a temperature of 50 keV (for instance), and with a 45-degree cone angular distribution around the perpendicular to the gradient like that shown in Fig (8). As an energetic electron enters from the left, another energetic electron would exit this boundary, to maintain neutrality. Similarly, when an energetic electron, due to collisions, leaves on the left, a background electron comes back in. These energetic electrons would move through the computational region and collide with the background electrons and ions. The right-hand boundary might have a background density of about 0.3nc When the energetic electrons leave on the right, they are replaced with an electron entering on the right, but they would not be cold, but would have a distribution with a temperature of the assumed temperature of the energetic electrons.

However, these collisions could not be binary. When the electron scatters by 90 degrees, it does not do so by single binary collisions. If it did, they would rapidly heat the background plasma. Instead it is described by many small angle scatterings. An earlier work of the author and his colleagues showed how to model this in a PIC simulation of electrons (46), and the method there could easily be extended to the problem at hand.

To this author, it seems that one could formulate the problem in this way as an initial value problem and run the code until a steady state is reached. The calculated distribution on the right-hand side, would be the boundary condition to impose on the Fokker Planck equation. It would be interesting to see what this steady state distribution function would look like as a function of Z. For high Z, where the dominant energy conserving collision are very large compared to the frictional force, the drifted Maxwellian distribution would probably be pretty good. However, when the Z gets to be 1, the energy conserving collisions would only be double the collisions causing friction, and there might be considerable reduction of the energetic flux, balanced by heating in the localized region.

E. Monte Carlo simulations of the electron orbits can play a role.

While the author does not see such Monte Carlo orbit calculations, as shown in Fig 10, as playing a role as a subroutine called at each time step of a rad hydro code, that does not mean it cannot play some role. For instance, to check on the Fokker Planck simulations at, let's say, the time of maximum incident energetic electron heat flux, one might use a Monte Carlo orbit simulation. Surely the simulation would need many thousands of electrons to sufficiently populate the energy and angular distribution and account for the spherical symmetry. The author has no experience with these codes and does not know whether they are equipped to calculate density from many orbits. If the codes can do this, fine; if they cannot, it almost certainly would be possible to incorporate such a capability. Using these codes would also be a check, on just a single time step, on the validity of the Fokker Planck, which is used every time step,

F. Let's reconsider fusion breeding.

Since the fusion reaction produces only a single neutron, how can it sustain a fusion economy which where some neutrons will be lost, and some tritium produced will be lost or decay (it has a half-life of 12 years). Hence if tritium can only be generated by a fusion neutron, and nothing else, the tritium which the project started out with will itself have a half-life, and a fusion economy will be impossible to achieve.

Fortunately, there are other sources. Since the fusion neutron has a high energy, 14 MeV, as opposed to ~ 2 MeV from fission reactions, this energetic neutron can produce other neutrons by spallation if the target is a neutron multiplier like beryllium, lead, 238U…. Fusion designers realize this and plan to place these elements in the blankets so that each fusion neutron produces let's say 1.5 tritons (31, 47,48). This is enough to make up for the various losses, so that the number of fusion reactions can grow exponentially in time.

However, this author believes that it is possible to do better, much better. First, once we have all the fusion reactors we need, what will we do with the extra half neutron, just throw it away? What a terrible, wasteful idea, especially where mined fissile material (235U) may well be in short supply in coming years. Why not just use it to breed fissile material from fertile material, say 233U from thorium? The author has studied and advocated this over decades, (31,49). Fission breeders cannot make up the deficit, it takes 2 fission breeders at maximum breeding rate to fuel a single thermal fission reactor (like the light water reactor) of equal power. Uranium from the seas may or may not be able to make up the slack, it is hardly a panacea (50, 51). But a single fusion breeder can fuel 5 or 10 thermal reactors! The reason is that fission is energy rich and neutron poor, while fusion is neutron rich and energy poor, a perfect match (31,49)

Forty or more years ago, preliminary calculations were done on potential blankets, indicating that these could be designed to breed both tritium as required, and also 233U (52-54). To this author's knowledge, these calculations have not been followed up, certainly not by groups eager to show that there is something there. Over nearly the entire life of the fusion project, fusion breeding has been the ugly duckling, anybody who advocated it, would promptly get his support pulled. This author has written several scientific papers on it, *none* of which acknowledged a sponsor. It is time to reverse this and let fusion breeding become the beautiful swan, *which it is* (31)*!* It is time to redo these calculations, with the psychology that they will succeed, not with the psychology that they are there to eliminate fusion breeding. They do not quit and loudly announce failure, because the first attempt does not produce quite enough because of a diagnostic port somewhere on the blanket. Certainly, the simplest calculations (52-54), show that a 14 MeV neutron impinging on the proper material can produce more than enough tritons and 233U.

## X Conclusions

In this section the author briefly discusses his optimism regarding laser fusion. As stated in the introduction, he sees three tall poles supporting this optimism. The tallest, of course is the LLNL experiments on getting Q~2.5 with its NIF laser with light energy of ~2 MJ. . This was accomplished with an indirect drive configuration, where the laser produced first X-rays, and these imploded the target. This has an advantage in that it borrows as much as possible the nuclear weapons knowledge. It has several disadvantages in that only a small fraction of the laser light energy, ~10-15% is used to implode the target. Not only that, but it also brings the relatively large heavy metal hohlraum into the picture. If one is interested in energy rather than nuclear simulation, one not only needs it for each shot, but one also must get rid of it before the next shot.

The second tall pole is the URLLE work. The alternative is to do away with the hohlraum and have the full power of the laser focused on each target. One would like the laser to have as short a wavelength as possible; most of the work has been done with frequency tripled light with a wavelength of 335 nm. The question, of course is can ultraviolet light produced the required implosion as well as X-Rays. Many people are skeptical. However, URLLE did a wonderful series of scaled experiments with their 30 kJ laser OMEGA, indicating, that at least within the confines of their experiment, they were successful in producing the scaled ultraviolet driven implosion that a 2 MJ ultraviolet laser would be expected to produce, and which NIF did produce using X-Rays to implode.

The third tall pole is the NRL work on developing excimer lasers, first KrF, and later ArF. These not only have shorter wavelength but also have several other capabilities which are advantageous for energy production. Not only is their efficiency likely higher, but the active material is a fast-flowing gas, not a pile of laser glass which must be cooled down for hours between shots. It likely excimers lasers have higher average power capability; they also have several other advantages.

Of course, none of these 'tall poles' are conclusive proof that laser fusion with a direct drive configuration will work. There are other complications. For, instance the ultraviolet laser used for fusion, might produce energetic electrons on the surface of the target, which most likely did not occur in the URLLE experiments, and which might cause destructive preheat of the fuel. That has motivated this work, and while the author certainly cannot claim a triumph, he believes that he has found significant reason for optimism as shown in Figs (22 and 25).

Hence the author feels that using an excimer laser with an energy of 1-5 MJ is the optimal approach to fusion now, and the proper strategy for our political bosses is to implement it (31,32). While there are many roadblocks and speed bumps along the way, not all foreseeable currently; he is convinced that this best way to go, given what we know now, not only about laser fusion, but also what we know about magnetic fusion.

Furthermore, let's degrade the NRL estimates. Let's say the laser has an efficiency of 'only' 7%, and the fusion process has a gain of 'only' 50. It is simple to see that with these numbers, this is not viable for generating electricity. However pure fusion is not the only possible outcome. As this author has argued for years (31), fusion can be a very prolific generator of fissile fuel for thermal nuclear reactors. The author has called this 'fusion breeding'. The basic reason for this is that as nuclear reactions go, fusion is neutron rich, but energy poor; while fission reactions are energy rich and neutron poor; a perfect match.

The ‘degraded’ laser fusion just specified would work fine for fusion breeding (31). Now let’s see where we are. With perhaps with only 10% of the laser energy imploding the target, the LLNL project has produced a gain of 2.5. If an ultraviolet experiment works as well, when *all* the laser energy could hit the target, as URLLE results indicate, and as the NRL group had assumed, the LLNL result would have demonstrated a gain of 25! Yet a gain of ‘only’ 50 would be sufficient for fusion breeding. As far as required gain for the fusion process goes, *we may already be halfway there!*

This concludes with arguments against what some have proposed. Some, worrying about the inefficiency of the laser, have proposed using a more efficient heavy ion accelerator instead of a laser to drive the implosion. First, no such accelerators exist, and it is not exactly a simple task to build one. The Russian effort spoke of a 7 km long accelerator, supplemented with 7 large storage rings (49,55). This does not sound simple or cheap.

Others, worried about the need to breed tritium fuel, have suggested using the DD reaction. Even at Megavolt temperatures, the DD reaction rate is at least two orders of magnitude less than the DT reaction rate. At conventional fusion temperatures of 5-20 keV, the situation is even worse. The need to breed tritium is indeed a difficult problem, largely ignored in the fusion world. To be viable for fusion, the blanket must provide considerably more than one T that the fusion gives. Neutron multipliers must supply the rest, so that the number of fusion reactors can grow exponentially in time.

This author feels that jumping into either of these projects is like telling our sponsors that after three quarters of a century of fusion research, we suddenly realized that we still need another half century or more, just to get to the starting gate! Surely it is extremely unlikely in this circumstance, that our sponsors would continue to sign the checks. Just how much patience do we think they have?

Right now, it is likely that we are finally climbing out of the bottomless pit that fusion has been stuck in for 60 years. Let’s not now jump into one or two other bottomless pits! Now it is like third down and 20 yards to the goal line. Let’s be Tom Brady and make that damn touchdown; we can do this! Let’s not say we will only play if the goal line is moved back a kilometer, but we need 100 more downs. Rather, let’s attempt to exploit the successes we have already achieved. Let’s build a fusion reactor, and/or a fusion breeder using a 1-5MJ laser, which we already have, work on improving its efficiency to 10%, or even 7%, work on other technical aspects of the reactor, tracking the target, manufacturing targets, injecting them, blanket issues … and of course, work on getting a fusion process with a gain of at least 50. As regards the latter, we may already be halfway there. These are big issues, which will still take decades to resolve. This author feels that at this early stage, they would be dealt with best on the national lab scale, hopefully a new national lab dedicated to laser fusion for energy.

Acknowledgement: This work was not supported by any agency, public or private. The author thanks Steven Obenschain, formerly the leader of the NRL project, currently the leader of the Laser Fusion X company for supporting the author's work in this area.